\def\bfseries{\fontseries\bfdefault\selectfont\boldmath}
\def\itshape{\fontshape\itdefault\selectfont\let\mathrm=\mathit}

\pdfoutput=1
\RequirePackage{lineno}
\documentclass[twocolumn,showpacs,preprintnumbers,amsmath,amssymb]{revtex4}

\newcommand{\mtw}{$m_{T}^{W}$}
\newcommand{\met}{$E_{T}^{Miss}$}

\newcommand{\tev}{TeV}
\newcommand{\gev}{GeV}
\newcommand{\GeV}{GeV}
\newcommand{\etaa}{$|\eta|$}

\newcommand{\pT}{$p_T$}

\newcommand{\ttbar}{$t\bar{t}$}

\usepackage{multirow}
\usepackage{lineno}
\usepackage{atlasphysics}
\usepackage{subfigure}
\usepackage{graphicx}
\usepackage{dcolumn} 
\usepackage[hyperindex,breaklinks]{hyperref} 

\usepackage{preprintcover}  
\PreprintCoverPaperTitle{Search for a multi-Higgs-boson cascade in $W^+W^-b\bar{b}$ events \\
 with the ATLAS detector in $pp$ collisions at $\sqrt{s}$ = 8 \TeV}
\PreprintIdNumber{CERN-PH-EP-2013-173}  
\PreprintCoverAbstract{A search is presented for new particles in an extension to the Standard
 Model that includes a heavy Higgs boson ($H^0$), an intermediate charged Higgs
 boson pair ($H^\pm$), and a light Higgs boson ($h^0$). The analysis searches for events 
 involving the production of a single heavy neutral Higgs boson which decays to the charged Higgs boson and a $W$ boson, 
where the charged Higgs boson subsequently decays into a $W$ boson and
the lightest neutral Higgs boson decaying to a bottom--antibottom-quark pair. 
Such a cascade results in a $W$-boson pair and a bottom--antibottom-quark 
 pair in the final state. Events with exactly one lepton, missing transverse momentum, and at least four jets are 
selected from a data sample corresponding to an integrated luminosity of 20.3 fb$^{-1}$, collected 
by the ATLAS detector in proton--proton collisions at  $\sqrt{s}=8$
TeV at the LHC. The data are found to be consistent with Standard Model
predictions, and 95\% confidence-level upper limits are set on the
product of cross section and 
 branching ratio. These limits range from 0.065 pb to 43 pb as a function 
 of $H^0$ and $H^\pm$ masses, with $m_{h^{0}}$ fixed at \mbox{125 GeV}.}
\PreprintJournalName{PRD}  

\begin{document}

\begin{abstract}
 A search is presented for new particles in an extension to the Standard
 Model that includes a heavy Higgs boson ($H^0$), an intermediate charged Higgs
 boson pair ($H^\pm$), and a light Higgs boson ($h^0$). The analysis searches for events 
 involving the production of a single heavy neutral Higgs boson which decays to the charged Higgs boson and a $W$ boson, 
where the charged Higgs boson subsequently decays into a $W$ boson and
the lightest neutral Higgs boson decaying to a bottom--antibottom-quark pair. 
Such a cascade results in a $W$-boson pair and a bottom--antibottom-quark 
 pair in the final state. Events with exactly one lepton, missing transverse momentum, and at least four jets are 
selected from a data sample corresponding to an integrated luminosity of 20.3 fb$^{-1}$, collected 
by the ATLAS detector in proton--proton collisions at  $\sqrt{s}=8$
TeV at the LHC. The data are found to be consistent with Standard Model
predictions, and 95\% confidence-level upper limits are set on the
product of cross section and 
 branching ratio. These limits range from 0.065 pb to 43 pb as a function 
 of $H^0$ and $H^\pm$ masses, with $m_{h^{0}}$ fixed at \mbox{125 GeV}.
\end{abstract}

\title{Search for a multi-Higgs-boson cascade in $W^+W^-b\bar{b}$ events \\
 with the ATLAS detector in $pp$ collisions at $\sqrt{s}$ = 8 \TeV}

\author{The ATLAS Collaboration}
 

\maketitle


\section{Introduction}
\label{sec:introduction}
Recently, a Higgs boson has been discovered by the
ATLAS~\cite{atlashiggs} and CMS~\cite{cmshiggs} collaborations with a mass of approximately 125 GeV. 
This observation has been supported by complementary evidence from the CDF and D0 collaborations~\cite{cdfd0higgs}.
The study of such a boson, responsible for breaking
electroweak symmetry  in the Standard Model (SM),
is one of the major objectives of experimental high-energy physics. 
A vital question is whether this state is in fact the Higgs
boson of the SM, or part of an extended Higgs
sector (such as that of the minimal supersymmetric Standard Model,
MSSM \cite{mssm1,mssm2}), a composite Higgs boson \cite{comphiggs},
or a completely different particle with Higgs-like couplings
(such as a radion in warped extra dimensions \cite{radion1,radion2}
or a dilaton \cite{dilatonHiggs}).

This Letter reports a search for particles in an extension to the SM that includes heavier Higgs bosons in addition to a light neutral Higgs boson, $h^0$, with mass $m_{h^0}=125$ GeV. Rather than assuming 
a particular theoretical  model, this analysis follows a simplified model 
approach by searching for a specific multi-Higgs-boson cascade topology~\cite{pheno}.
Many beyond-the-SM Higgs models introduce a second Higgs doublet. In addition to the $h^0$, such models contain a heavy charged Higgs boson pair
$H^\pm$ and a heavier neutral state $H^0$.  An additional pseudoscalar particle, $A$,
may also exist within the Two Higgs-Doublet Model (2HDM)~\cite{2hdm} framework, but this analysis assumes it to be too 
heavy to participate in the cascade decay considered here.

\begin{figure}[h]
\begin{center}
\includegraphics[width=3in]{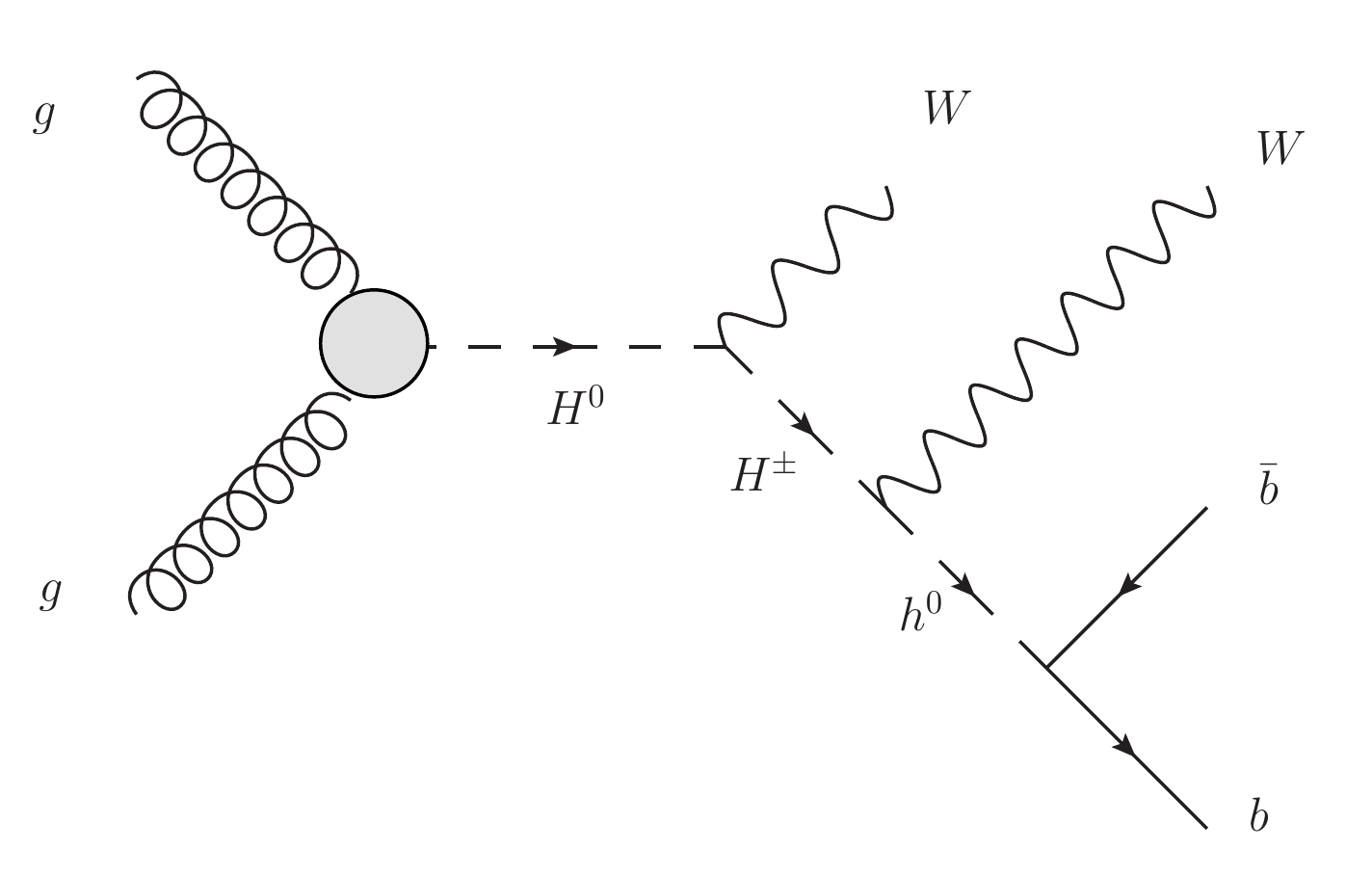}
\caption{Diagram showing the Higgs-boson cascade  $gg \to H^0 \to
  W^{\mp}H^\pm  \to  W^\mp W^\pm h^0 \to  W^{\mp}W^\pm b\bar{b}$.}
\label{fig:diag}
\end{center}
\end{figure}

This letter reports the first search at the LHC for new particles in
the final state $W^\pm W^\mp b\bar{b}$, via the process $gg \to H^0$ 
followed by the cascade, 
 $H^0 \to
  W^{\mp}H^\pm  \to  W^\mp W^\pm h^0 \to  W^{\mp}W^\pm b\bar{b}$,
as illustrated in Fig.~\ref{fig:diag}. Other production modes, such as associated production or vector-boson fusion lead to different final states and are not considered here.
  The $W^\pm W^\mp b\bar{b}$ final
state also appears in top-quark pair production. In this search, one of the $W$ bosons
is assumed to decay to hadrons leading to jets and the other one decays to an electron plus a neutrino
($e$+jets) or a muon plus a neutrino ($\mu$+jets). The same final state has been used by
CDF in a similar search for Higgs-boson cascades~\cite{cdfprl}.
Other related searches have been performed for charged
Higgs bosons in top-quark decays $t\rightarrow H^+
b$~\cite{chiggs0,chiggs1,chiggs2,chiggs3,chiggs4,chiggs5}. Boosted
decision trees (BDTs) are used to
distinguish the Higgs-boson cascade events from the predominantly \ttbar background.

\section{ATLAS detector and data sample}
\label{sec:detector}
The ATLAS experiment~\cite{atlasdetector} at the LHC is a multi-purpose particle physics detector with 
approximately forward-backward symmetric cylindrical geometry~\cite{atlascoord}. It consists of an 
inner tracking detector surrounded by a thin superconducting solenoid,
electromagnetic and hadronic calorimeters, and a muon spectrometer
incorporating three large superconducting toroid magnet assemblies. 

The data used in this analysis were collected during 2012 from $pp$
collisions at a centre-of-mass energy of 8 TeV using triggers designed
to select high transverse momentum ($\pT$~\cite{atlascoord}) electrons or muons. The data sample corresponds 
to an integrated luminosity of 20.3~fb$^{-1}$.

\section{Signal and background simulation}
\label{sec:mcsimulation}
The production of  $H^0$ bosons via gluon fusion with $m_{H^0} = 325$--$1025$~GeV and subsequent decays $H^0\to W^\mp H^\pm$ with
$m_{H^\pm}=225$--$925$~GeV and $H^\pm\to W^\pm h^0$ with
$m_{h^0}=125$~GeV, is modelled using the MADGRAPH~\cite{madgraph} Monte Carlo (MC) event generator with an effective vertex to model the fermion loop and a narrow natural width of 50 MeV.  Additional radiation, hadronization, and
showering are described by PYTHIA v6.4~\cite{SJO-0601}. Thirty-six different mass
pairs are tested for the Higgs-boson cascade signal within the above
$m_{H^\pm}$ and $m_{H^0}$ mass ranges.

The dominant SM background to this signature is top-quark
pair production. This background is modelled using simulated events from the 
{\sc MC@NLO} v4.01~\cite{FRI-0201} event generator with
the CT10~\cite{ttbar_xsec_10} parton distribution functions (PDFs). The parton shower and the underlying event simulation are performed with {\sc HERWIG}
v6.520~\cite{COR-0001} and {\sc JIMMY} v4.31~\cite{Butterworth:1996zw},
respectively, using the AUET2 tune~\cite{PUB-2011-008}. The $t\bar{t}$
cross section for 
$pp$ collisions at a centre-of-mass energy of $\sqrt{s} = 8 \tev$
is assumed to be $\sigma_{t\bar{t}}= 253^{+13}_{-15}$~pb for a top-quark mass of $172.5 \gev$.  It has been
calculated at next-to-next-to-leading order (NNLO) in QCD including resummation of 
next-to-next-to-leading-logarithmic soft gluon terms with 
Top++2.0~\cite{ttbar_xsec_1,ttbar_xsec_2,ttbar_xsec_3,ttbar_xsec_4,ttbar_xsec_5,ttbar_xsec_6}. 
The PDF and $\alpha_s$ uncertainties are calculated
using the PDF4LHC prescription~\cite{ttbar_xsec_7} with the 
68\% confidence level (C.L.) of the MSTW2008 NNLO~\cite{ttbar_xsec_8,ttbar_xsec_9},
CT10 NNLO ~\cite{ttbar_xsec_10,ttbar_xsec_11} and NNPDF2.3 5f FFN ~\cite{ttbar_xsec_12} PDF sets, added in quadrature
to obtain the normalization and factorization scale uncertainties. Additional \ttbar\ samples used to estimate various systematic effects
are generated with {\sc POWHEG}~\cite{FRI-0701,powheg2,powheg3} interfaced to {\sc HERWIG}/{\sc JIMMY},
{\sc POWHEG} interfaced to {\sc PYTHIA}, and
{\sc AcerMC} v3.8~\cite{KER-0401} interfaced to
{\sc PYTHIA}. The \ttbar modelling is also checked with samples generated by {\sc ALPGEN}~\cite{MAN-0301}
interfaced with {\sc HERWIG}.

Other backgrounds are expected to originate from vector-boson production with
associated jets ($W$-boson+jets and $Z$-boson$/\gamma^*$+jets), as
well as single top-quark, diboson ($WW$,
$WZ$, $ZZ$), and multi-jet production.  All background predictions, except that for
multi-jet production, are obtained from simulated events.

The $W/Z$-boson+jets contribution is simulated using {\sc ALPGEN}
interfaced to {\sc HERWIG}/{\sc JIMMY}, and is normalized to NNLO theoretical 
cross sections~\cite{W_NNLO_1,W_NNLO_2}.
The contribution from single top-quark production is simulated using {\sc MC@NLO}
interfaced to {\sc HERWIG}/{\sc JIMMY} for the $s$-channel top-quark production and $Wt$ production, and with
{\sc AcerMC} interfaced to {\sc PYTHIA} for the $t$-channel, and normalized to
approximate NNLO theoretical cross sections~\cite{Kidonakis:2011wy,Kidonakis:2010tc,Kidonakis:2010ux}.  Finally,
diboson production is simulated with {\sc HERWIG} and normalized to 
NLO theoretical cross sections~\cite{Campbell:1999ah}.

All generated events are passed through the detailed ATLAS detector
simulation~\cite{Aad:2010ah} based on {\sc GEANT4} \cite{AGO-0301}, with the exception
of the additional samples used to account for systematic effects in
\ttbar production,  for which a parameterized simulation~\cite{Aad:2010ah} of the calorimeter
response is used. The events are then processed with the same reconstruction software as the data.
MC events are overlaid with additional minimum bias events generated
with {\sc PYTHIA} to simulate the effect of pile-up (additional $pp$
interactions in either the same or closeby bunch crossings as the primary interaction); the number of overlaid proton--proton interactions is chosen to match the distribution of the number of additional interactions observed in the data.

Multi-jet production may mimic the presence of a lepton, but the
contribution from these processes is found to be small.
It is estimated from the data by the matrix 
method~\cite{Aad:2010ey} in the $\mu$+jets and $e$+jets channels. 
The matrix method is a technique to estimate the number of events with a fake, 
isolated lepton in the signal selection, and uses \emph{loose} and \emph{tight} 
isolation definitions for leptons. The tight isolation definitions are those used in this analysis, and tight leptons are a subset of the loose leptons.
In a selection dominated by real leptons, the efficiency 
($\epsilon_{\textrm{real}}$) of a loose lepton to also pass the tight isolation requirements is measured. 
The rate ($\epsilon_{\textrm{fake}}$) of loose leptons passing the tight requirements is measured in a 
multi-jet-dominated selection. These rates, $\epsilon_{\textrm{real}}$ and $\epsilon_{\textrm{fake}}$, are
used to estimate the multi-jet contribution to the analysis selection.

\begin{figure*}[ht]
\begin{center}
\includegraphics[width=0.857\columnwidth]{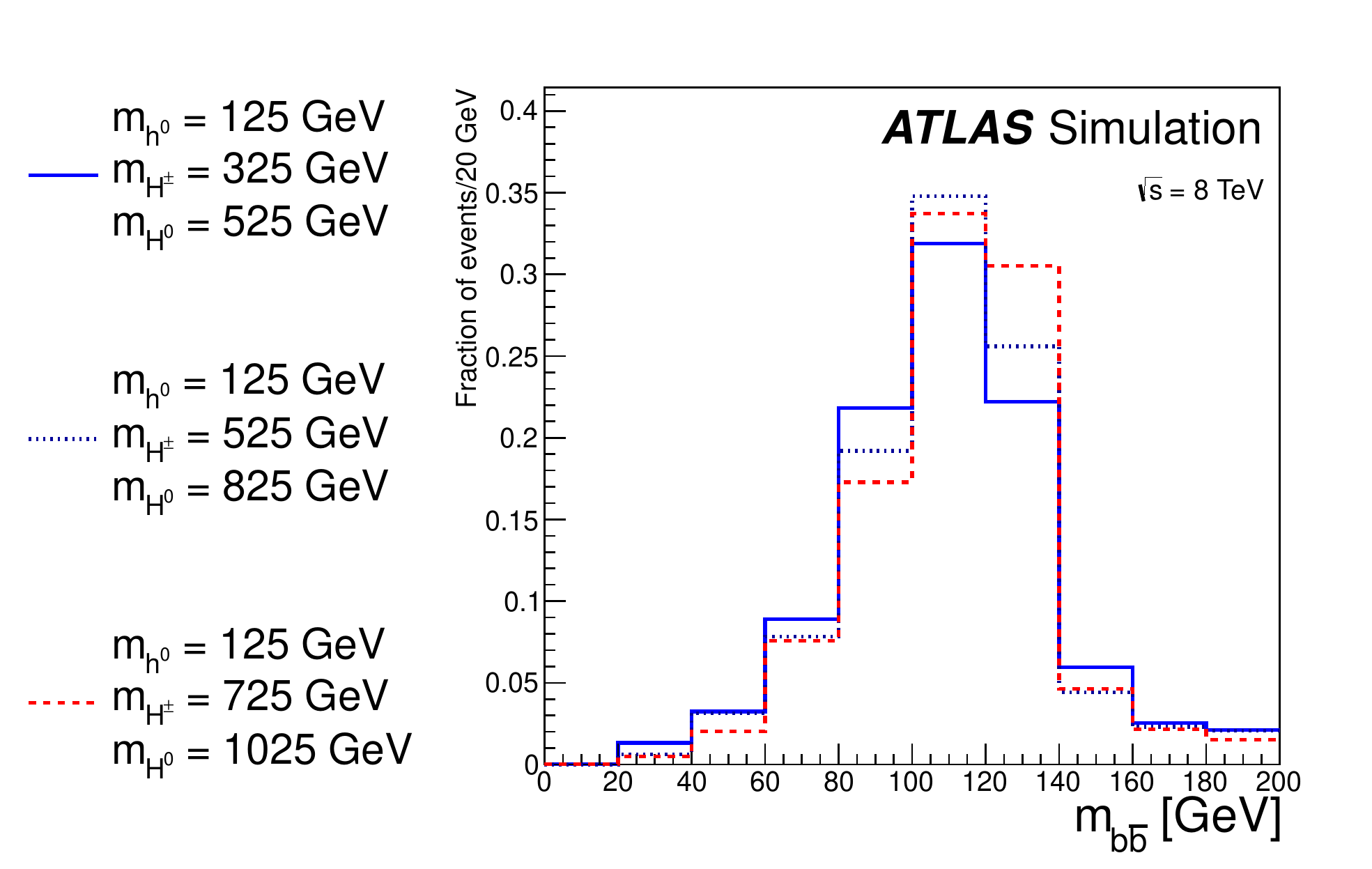}
\includegraphics[width=0.571\columnwidth]{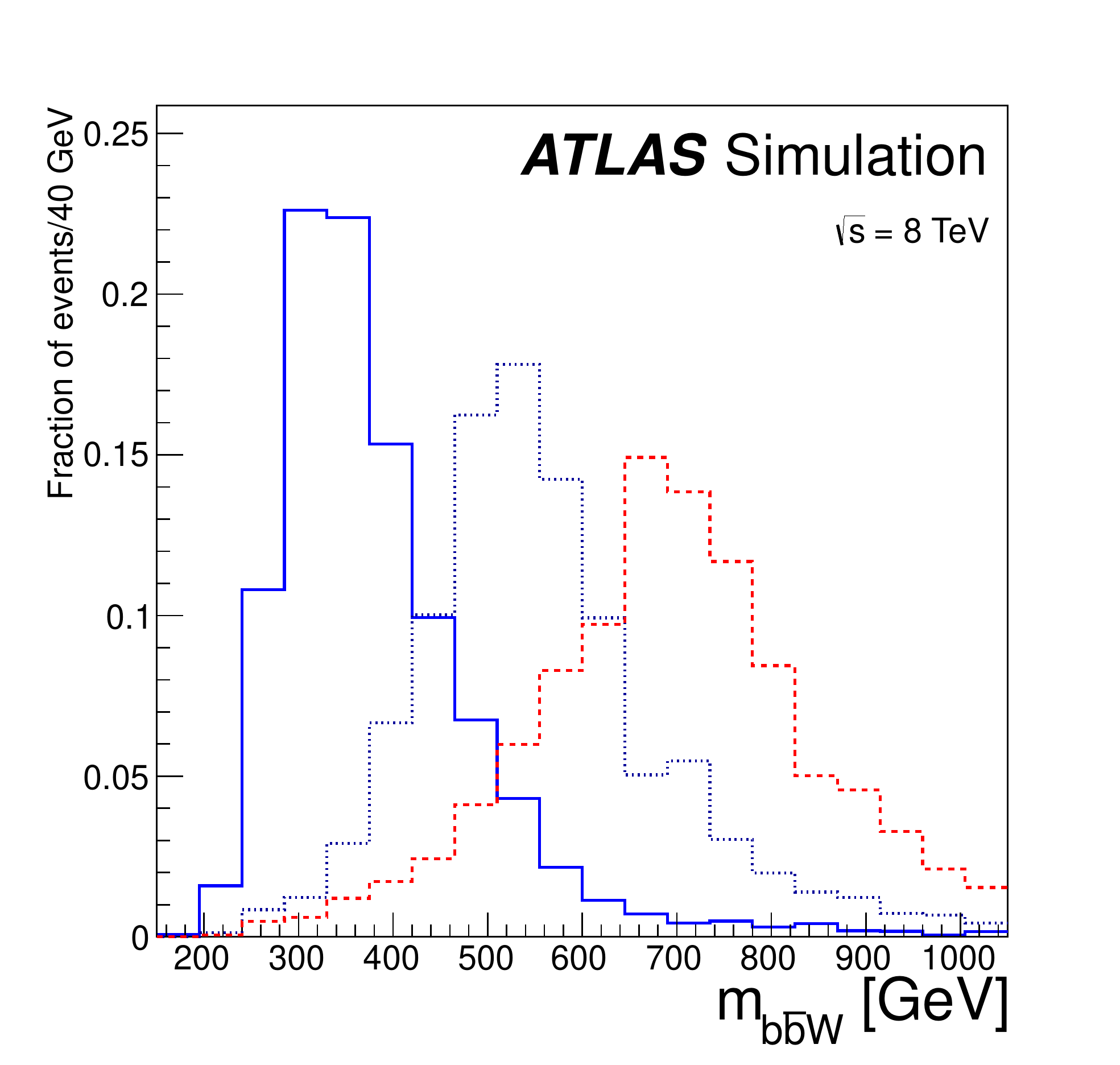}
\includegraphics[width=0.571\columnwidth]{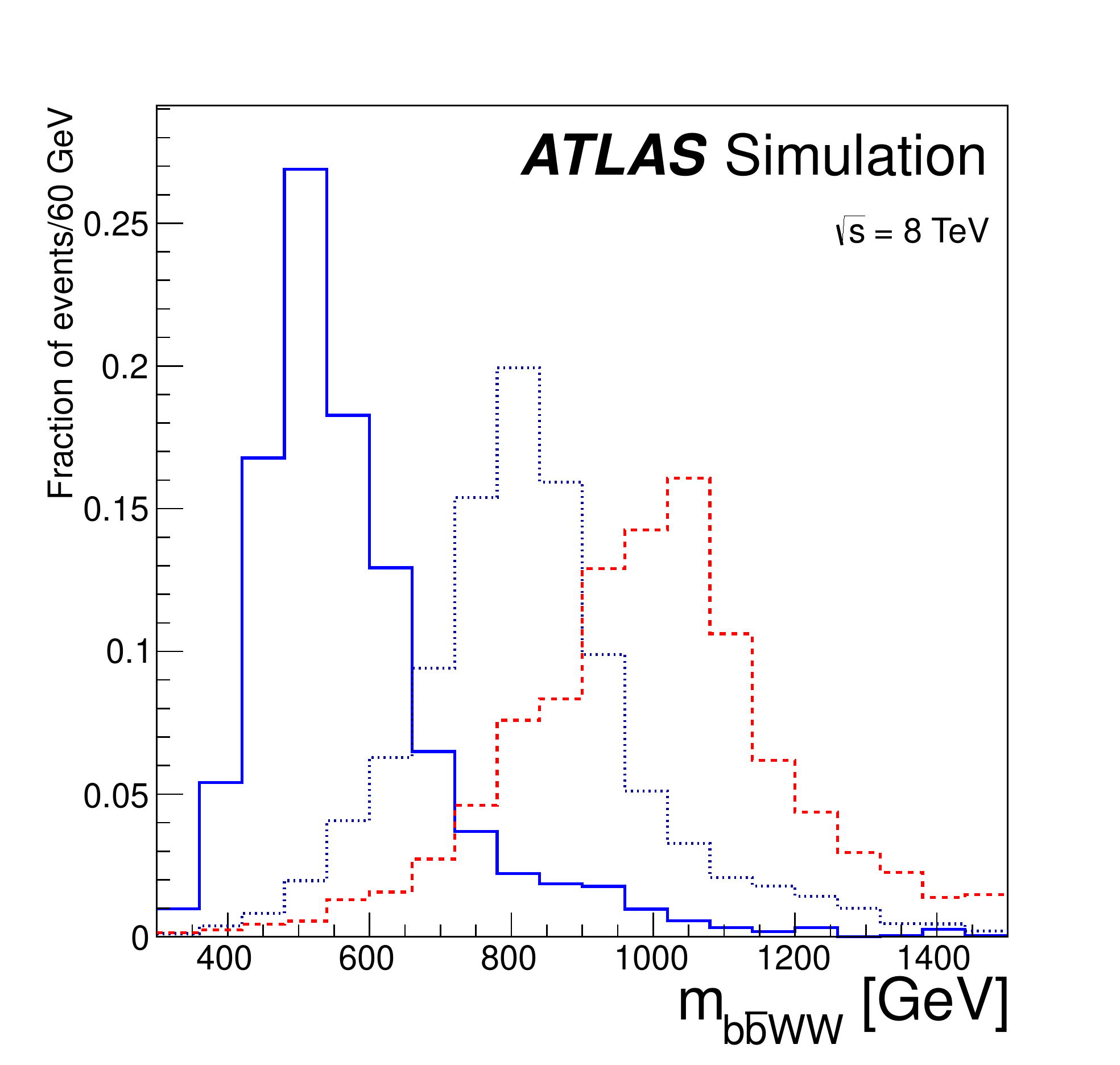}
\caption{Distributions of  reconstructed masses in simulation for the three Higgs bosons in the cascade;
the lightest Higgs boson, $h^{0}$ (left, as $m_{b\bar{b}}$), the charged Higgs boson,
$H^{\pm}$ (middle,  as $m_{b\bar{b}W}$), and the heavy Higgs boson, $H^{0}$ (right,  as $m_{b\bar{b}WW}$), 
shown for three example mass hypotheses.}
\label{fig:recon}
\end{center}
\end{figure*}

\section{Event selection}
\label{sec:object_def}
This analysis relies on the measurement of jets, electrons, muons and
the missing transverse momentum (\met)~\cite{MET}. Since this analysis investigates a final state dominated by top-quark pair production, 
a  selection similar to the top-quark cross-section measurement by the ATLAS collaboration~\cite{tt_xsec} is used.

Jets are reconstructed using the anti-$k_t$ algorithm~\cite{Cacciari:2008gp} with a radius parameter
$R=0.4$, and are calibrated at the energy cluster level~\cite{LC} to compensate for differing calorimeter response
to hadronic and electromagnetic showers. A correction for pile-up
is applied to the jet energy~\cite{jet_area}. Jets are required to have $p_T>$ 25 GeV and $|\eta|<$ 2.5.  
Jets from additional $pp$ interactions are suppressed by
requiring the jet vertex fraction (JVF) to be larger than 0.5 for jets with $\pT<$ 50 GeV
and $|\eta|<$2.4. The JVF variable is defined as the transverse momentum weighted fraction 
of tracks associated with the 
jet that are compatible with originating from the primary vertex.
The primary vertex is defined as the vertex with the largest $\sum p_T^2$ of associated tracks.
Jets are $b$-tagged (identified as the product of a $b$-quark) using the MV1 tagger~\cite{btag}, which combines several
tagging algorithms~\cite{jet_algos} using an artificial neural network. 
A 70\% tagging efficiency is achieved in identifying 
$b$-jets with $\pT>$ 20 GeV and $|\eta|<$ 2.5 in simulated $t\bar{t}$ events, 
while the light-jet rejection factor is 130. Additional corrections to
the tagging efficiency and mistagging rate are derived from data and applied to all simulated samples~\cite{btag, btag1, ctag, mistag}.

Electrons are identified~\cite{el_reco} as energy clusters in the electromagnetic calorimeter
matched to reconstructed tracks in the inner detector. Selected electrons are required
to pass stringent selection requirements that provide good discrimination between isolated electrons and jets. Isolation requirements are imposed
in cones  of calorimeter energy deposits ($\Delta R(e,{\textrm{deposit}})<0.2$) and inner-detector tracks ($\Delta R(e,{\textrm{track}})<0.3$) around the electrons direction where $\Delta R= \sqrt{(\Delta\eta)^2 + (\Delta\phi)^2}$. The calorimeter isolation is corrected
for leakage of the energy of the electron into the isolation cone and for energy deposits from pile-up events. Both the calorimeter and the inner detector isolation requirements are
chosen to give 90\%
efficiency. Selected electrons are required to have transverse momentum $p_{\textrm{T}}>25$ \GeV and
pseudorapidity in the range $|\eta|<2.47$, excluding the calorimeter barrel/end-cap transition region $1.37<|\eta|<1.52$.  

Muons are reconstructed~\cite{mu_reco} using information from the muon spectrometer and the 
inner detector and are required to fulfil isolation requirements. 
Muons are required to have transverse momentum $\pT>25$ \GeV and  
$|\eta|<2.5$. The isolation variable~\cite{tt_res,miniiso} for muons is defined as 
$I_{\mu} = \sum \pT^{track}/\pT^{\mu}$, where the sum runs
over all tracks (except the one matched to the muon) that pass quality requirements and
have $\pT^{track}>$ 1 GeV and $\Delta R(\mu,track)<$ 10
GeV$/\pT^{\mu}$. Muons with
$I_{\mu}<$ 0.05 are selected.

The transverse momentum of neutrinos is inferred from the magnitude of the missing transverse 
momentum  in the event.
The missing transverse momentum is constructed from the negative
vector sum of the reconstructed jets, the topological calorimeter energy deposits outside of jets, and the muon
momenta, all projected onto the transverse plane.

Overlapping  objects are subject to a removal procedure. 
The  jet closest to a selected electron is removed, if it is within
$\Delta R (e, \textrm{jet})<$ 0.2. Electrons with $\Delta R (e,
\textrm{jet})<$ 0.4 to any remaining jets and
muons with  $\Delta R (\mu, \textrm{jet})<$ 0.4 between the muon and nearest jet are 
removed since their likely origin is hadron decays.

Events are selected using single-lepton triggers with $\pT$ thresholds
of 24 GeV or 36 GeV for muons
and 24 GeV or 60 GeV for electrons (the lower momentum triggers also apply
isolation requirements).
Events are required to have exactly one reconstructed isolated 
electron or muon matching the corresponding trigger object and a primary vertex reconstructed from at
least five tracks, each with $\pT>400$~MeV. 
At least four jets with $\pT> 25$ \GeV and $|\eta|<2.5$ are required, of which at least two 
must be identified as $b$-jets. Additional requirements to reduce the
multi-jet background are applied:
	\begin{itemize}
     \item in the $e$+jets channel: $\met>30$ \GeV and \mbox{\mtw $> 30$ \GeV},
     \item in the $\mu$+jets channel: 
    	\met$>20$ \GeV and \mbox{\mtw$+$\met$>60$ \GeV}.  
	\end{itemize}

\noindent
The transverse $W$-boson mass is defined as 
	\mbox{\mtw$=\sqrt{2 \pT^\ell \pT^{\nu} (1 - \cos(\phi^\ell - \phi^{\nu}))}$},
	where $p_T$ is the transverse momentum, $\phi$ is the azimuthal angle, and
	$\ell$ and $\nu$ refer to the charged lepton and the neutrino,
        respectively. Different requirements are used for the muon and
        electron channels due to
different levels of multi-jet background contamination. The signal pre-region (SPR) is defined to contain events that pass these requirements. Table~\ref{tab:SPR}
illustrates the expected yields of the background and the observed number of events in this region.

\section{Event Reconstruction and Multivariate Analysis}
\label{sec:massreco}

\subsection{Event Reconstruction}
The Higgs-boson cascade event reconstruction begins with identification of the
leptonically decaying $W$ boson. It is assumed that the missing transverse
momentum is due to the resulting neutrino. The neutrino pseudorapidity
is set to the value which results in an invariant mass of the lepton and neutrino
closest to the nominal $W$-boson mass~\cite{pdg}; in the case of
degenerate solutions, the smallest magnitude of
pseudorapidity is chosen. Next, the two $b$-tagged jets 
are used to reconstruct the lightest Higgs-boson candidate, $h^{0}$; if there are more than two $b$-tagged jets, the
two jets with the highest $b$-tagging scores~\cite{btag} are used. The hadronically decaying
$W$ boson is identified from the remaining jets as the pair with reconstructed dijet
mass closest to the nominal $W$-boson mass. 
The charged Higgs-boson candidate $H^\pm$ is constructed from the light $h^0$ and the
$W$-boson candidate which gives the larger value of $m_{H^\pm}$. The
heavy neutral Higgs-boson candidate $H^0$ is then formed as $b\bar{b}WW$. 
Fig.~\ref{fig:recon} illustrates the
reconstructed mass distributions for the $h^{0}$, $H^{\pm}$, and $H^0$ in
simulation for selected mass values.

Since the dominant background is top-quark pair production, the two
$b$-quarks and two $W$ bosons are combined in $Wb$ pairs to give top-quark candidates. The combination
  which minimizes the sum of the absolute value of their differences from the
  nominal top-quark mass~\cite{pdg} for both pairs is chosen.
 The
invariant masses of the top-quark candidates are useful to 
discriminate $t\bar{t}$ events from the Higgs-boson signal.
The masses ($m_t$, $m_{\bar{t}}$) of the two top-quark candidates and
the absolute values of their differences ($|m_t-m_{\bar{t}}|$) are calculated.

\subsection{Multivariate Analysis}
A multivariate analysis is performed to distinguish the Higgs-boson cascade from $t\bar{t}$ events. 
Several reconstructed kinematic quantities, including the invariant 
masses of the Higgs-boson candidates as described above, are used as inputs to a BDT classifier, 
provided in the TMVA~\cite{tmva} package. TMVA provides a ranking for the input variables, which is
derived by counting how often an input variable is used to split decision tree nodes,
and by weighting each split occurrence by the square of the gain in
signal-to-background separation it has achieved
and by the number of events in that node. Several combinations of input variables are tested in training 
the BDTs. The inputs for the BDTs are optimized for the best expected cross-section limits while avoiding 
overtraining, and the variable rankings of TMVA are used as heuristics in choosing the BDT inputs. Seven kinematic 
variables are chosen to achieve the best expected result across the entire signal mass grid:

\begin{itemize}
\item $m_{b\bar{b}}$, $m_{b\bar{b}W}$ and $m_{b\bar{b}WW}$, as
  described above;
\item $\Delta R(b,\bar{b})$, the angular distance between the pair of $b$-tagged 
jets used to reconstruct the light Higgs-boson candidate;
\item leptonic $m_t$, the top-quark mass reconstructed using the
  leptonically decaying $W$ boson;
\item hadronic $m_t$, the top-quark mass reconstructed using the
  hadronically decaying $W$ boson; 
\item $|m_t-m_{\bar{t}}|$.
\end{itemize}

For cascades originating from a high-mass Higgs boson, the reconstructed top-quark masses along with $m_{WWb\bar{b}}$ 
are the highest-ranked input variables. For the low-mass Higgs-boson cascades, $m_{b\bar{b}}$
and $\Delta R(b,\bar{b})$ have the highest rank.
Since the kinematics of the Higgs-boson cascade vary greatly with the masses of the
heavy and intermediate Higgs bosons, a different BDT is trained for each signal mass hypothesis. 

Only MC events that pass the SPR requirements are used in the training of the BDTs. Each 
BDT is constructed as a forest with 750 decision trees, and is trained against  simulated  background event samples. 
The stochastic gradient
 boosting method~\cite{tmva} is used to improve classification accuracy and its robustness against statistical fluctuations. Each BDT
is checked for overtraining with a statistically independent test sample.  

For each of the 36 signal mass points, a final threshold is chosen for its respective BDT output which gives the best expected sensitivity, measured using the same confidence-level calculations as applied to the data and described below. A counting experiment
is then performed using events that pass those BDT output thresholds. In this way, the 
BDT thresholds  divide the SPR into 36 non-orthogonal signal regions, one for each 
signal mass point. 

\begin{figure}[h]
\begin{center}
\includegraphics[width=0.85\columnwidth]{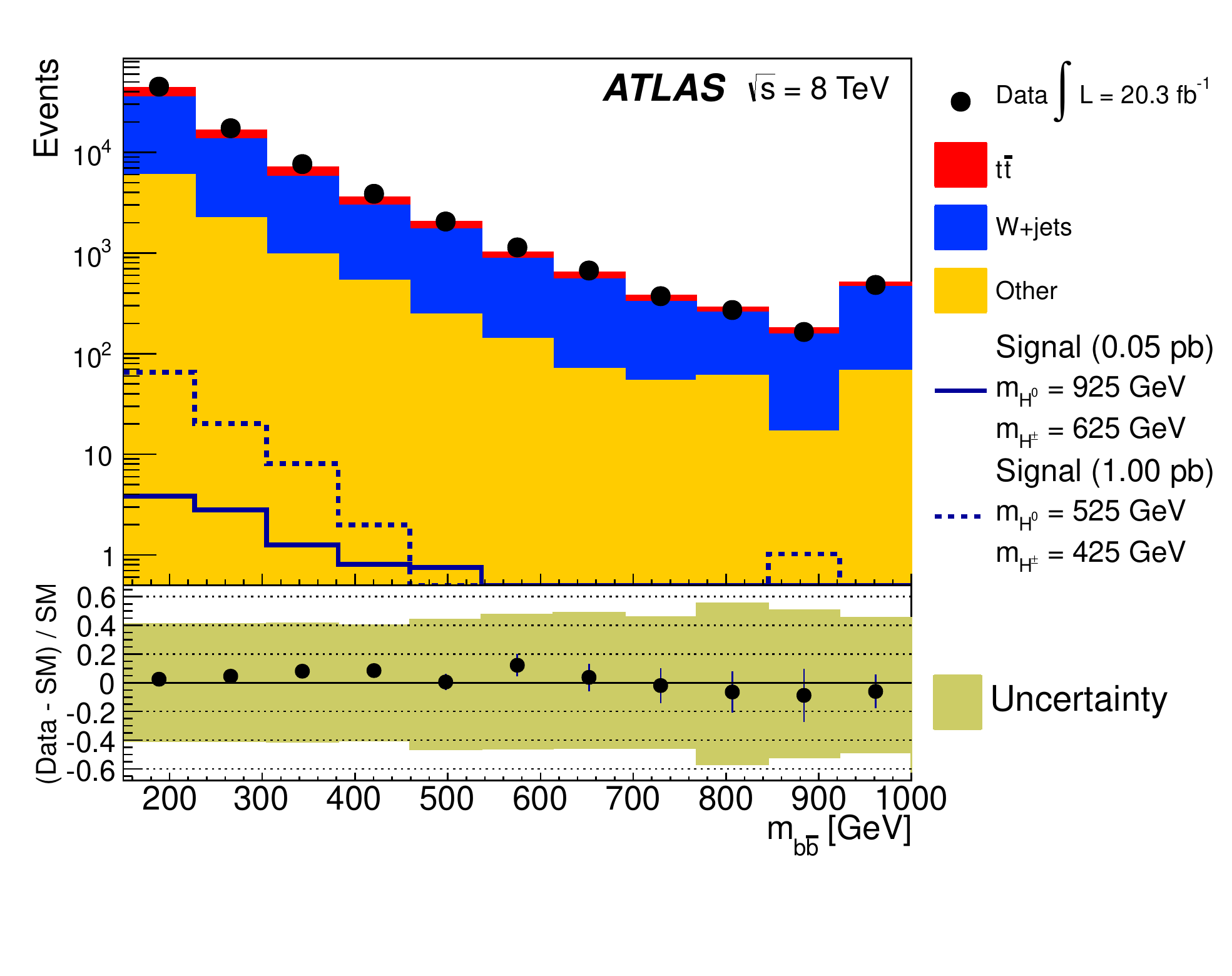}\\
\includegraphics[width=0.85\columnwidth]{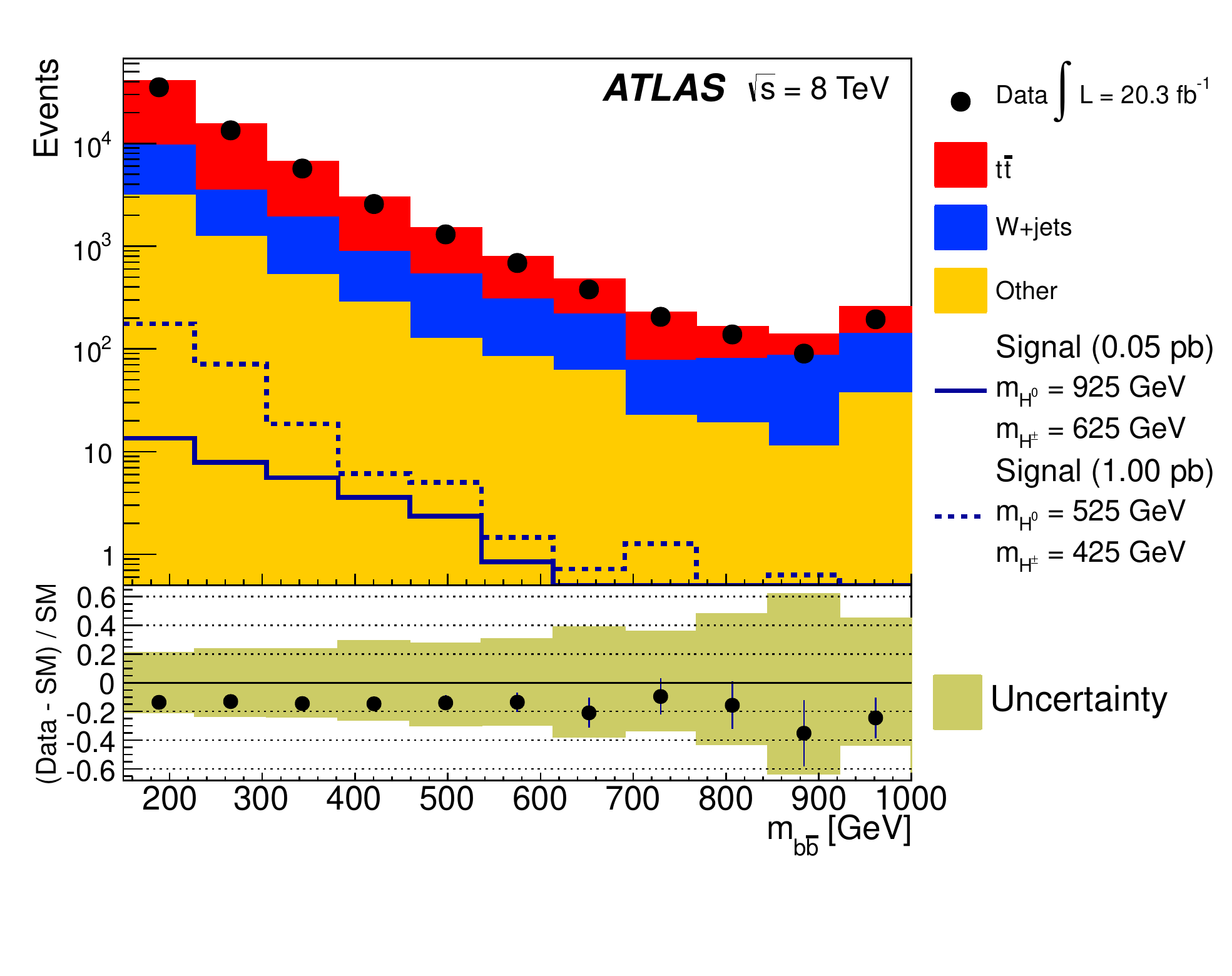}\\
\caption{Distributions of $m_{b\bar{b}}$ with uncertainties in the
  control regions CR1 (top) and CR2 (bottom). The data (black points) 
  are compared to the background model (stacked
  histogram).   In control regions with fewer than two 
$b$-tagged jets, the two jets with the highest $b$-tagging scores are
used. The final bin contains any overflow events. Two choices of signal hypotheses are also shown.}
\label{fig:control_cr2}
\end{center}
\end{figure}

\renewcommand{\arraystretch}{1.4} 
\begin{table} 
\caption{Expected background contributions with their total
  (systematic and statistical) uncertainties and the observed number
  of events with exactly one lepton and at least four jets, and in the
  SPR region, which additionally requires at least two $b$-tagged
  jets. In the table, contributions from processes with light-flavour
  (LF) $u,d,s$-quarks  and heavy-flavour (HF) $c,b$-quarks are distinguished.} 
\label{tab:SPR} 
\begin{center} 
\begin{tabular}{lrr} 
\hline 
\hline 
Source  &  $e/\mu$ + $\geq 4$ jets &  SPR Yields\\ 
\hline
\ttbar & $36.0^{+3.7}_{-3.8}$$\cdot 10^{4}$& $14.0^{+2.1}_{-2.0}$$\cdot 10^{4}$ \\
$W$-boson + jets LF & $16.0^{+8.2}_{-8.3}$$\cdot 10^{4}$& $6.0^{+4.2}_{-4.1}$$\cdot 10^{2}$ \\
$W$-boson + jets HF & $8.6^{+4.4}_{-4.4}$$\cdot 10^{4}$& $4.6^{+2.5}_{-2.4}$$\cdot 10^{3}$\\
$Z$-boson + jets LF & $26.0^{+6.3}_{-6.4}$$\cdot 10^{3}$& $11.0^{+8.3}_{-7.7}$$\cdot 10^{1}$\\
$Z$-boson + jets HF & $4.9^{+1.1}_{-1.0}$$\cdot 10^{3}$& $6.7^{+1.7}_{-1.6}$$\cdot 10^{2}$\\
Single top-quark & $16.0^{+2.0}_{-2.1}$$\cdot 10^{3}$& $46.0^{+7.6}_{-7.3}$$\cdot 10^{2}$\\
$WW, WZ, ZZ$ & $26.0^{+5.4}_{-5.5}$$\cdot 10^{2}$& $6.9^{+1.9}_{-2.0}$$\cdot 10^{1}$\\
Fake leptons & $1.8^{+1.8}_{-1.8}$$\cdot 10^{4}$& $8.6^{+8.6}_{-8.6}$$\cdot 10^{2}$\\  \hline 
Total & $68.0^{+14.0}_{-18.0}$$\cdot 10^{4}$& $15.1^{+2.2}_{-2.4}$$\cdot 10^{4}$\\
 \hline 
Observed  & 664876 &151123\\ 
 \hline 
 \hline 
 
 \end{tabular} 
 \end{center} 
 \end{table}

\section{Background Validation in Control Regions}
\label{sec:control}
The modelling of the SM backgrounds is validated in three
background-dominated control regions. The control regions retain the requirements of one lepton and at least four jets, and each region has additional requirements. In control regions with fewer than two 
$b$-tagged jets, the two jets with the highest $b$-tagging scores are used to reconstruct the lightest 
Higgs boson, $h^{0}$.  The following control regions are used:

\begin{itemize}

  \item Control Region 1 (CR1):  at least four jets,
    exactly one lepton and no $b$-tagged jets. This region
    validates primarily 
    the $W$-boson+jets modelling. This region is background-enriched relative
    to the hypothetical signal due to the $b$-tag veto.

  \item Control Region 2 (CR2):  at least four jets,
    exactly one lepton and exactly one $b$-tagged jet. This region
    validates primarily 
    the modelling of the $t\bar{t}$ background.  This background is fractionally larger, compared to a hypothetical signal, here than in the signal region due to the $b$-tagging cut, which preferentially selects the higher $p_{\textrm{T}}$ $b$-quarks from top-quark decay. Although a potential signal would not be absent in this control region, the different levels of signal and $t\bar{t}$ contributions allow a test of $t\bar{t}$ modelling by comparing levels of agreement between data and prediction in the signal and CR2 regions. 

  \item Control Region 3 (CR3):  at least four jets,
    exactly one lepton, at least two $b$-tagged jets, and
    $m_{b\bar{b}}>150$ GeV. This region  focuses primarily on validation the modelling of the
    $t\bar{t}$ background with kinematics similar to the hypothetical
    signal, but is background-enriched due to the $m_{b\bar{b}}>150$
    GeV requirement.

\end{itemize}

\begin{figure}[h]
\begin{center}
\includegraphics[width=0.85\columnwidth]{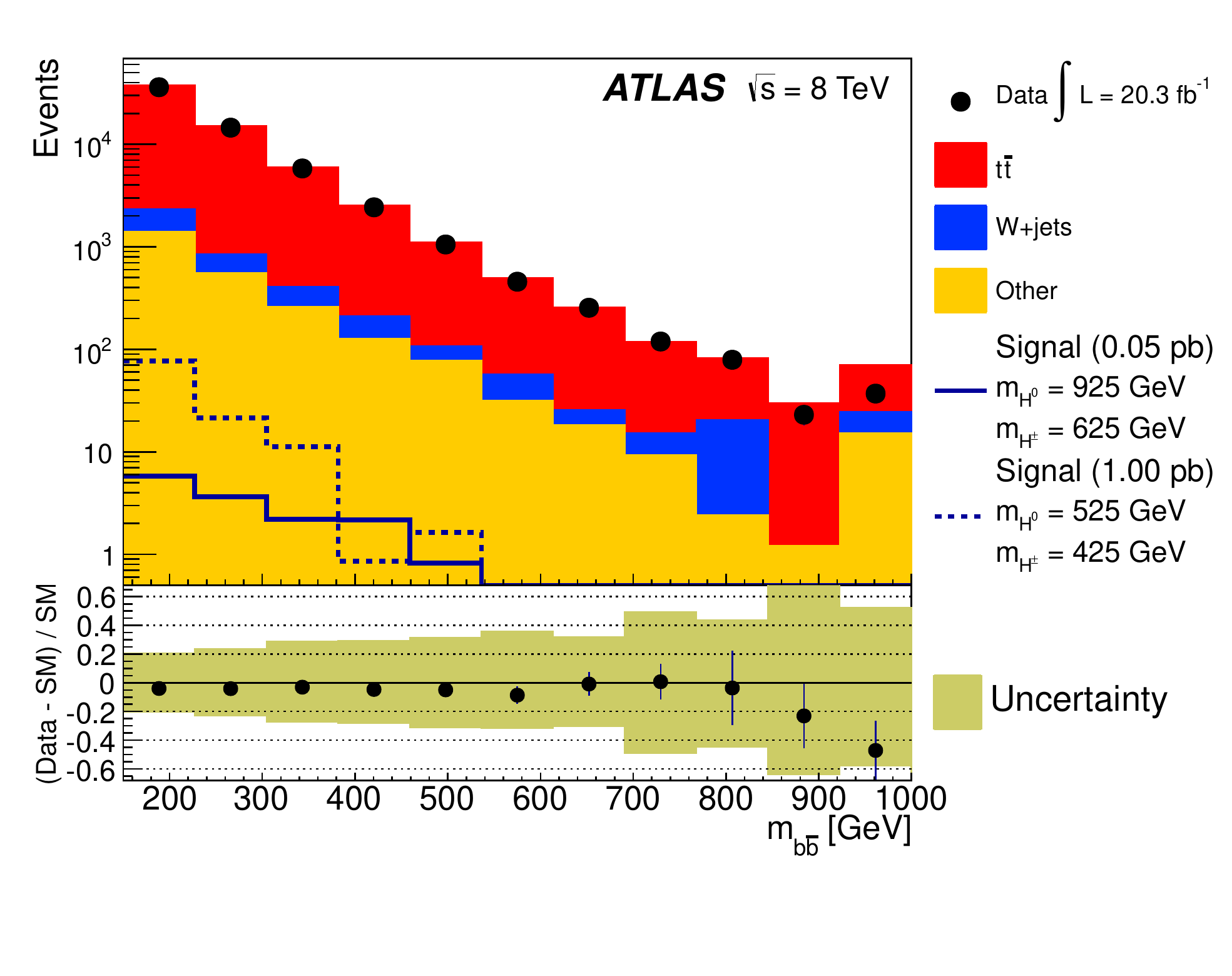}\\
\includegraphics[width=0.85\columnwidth]{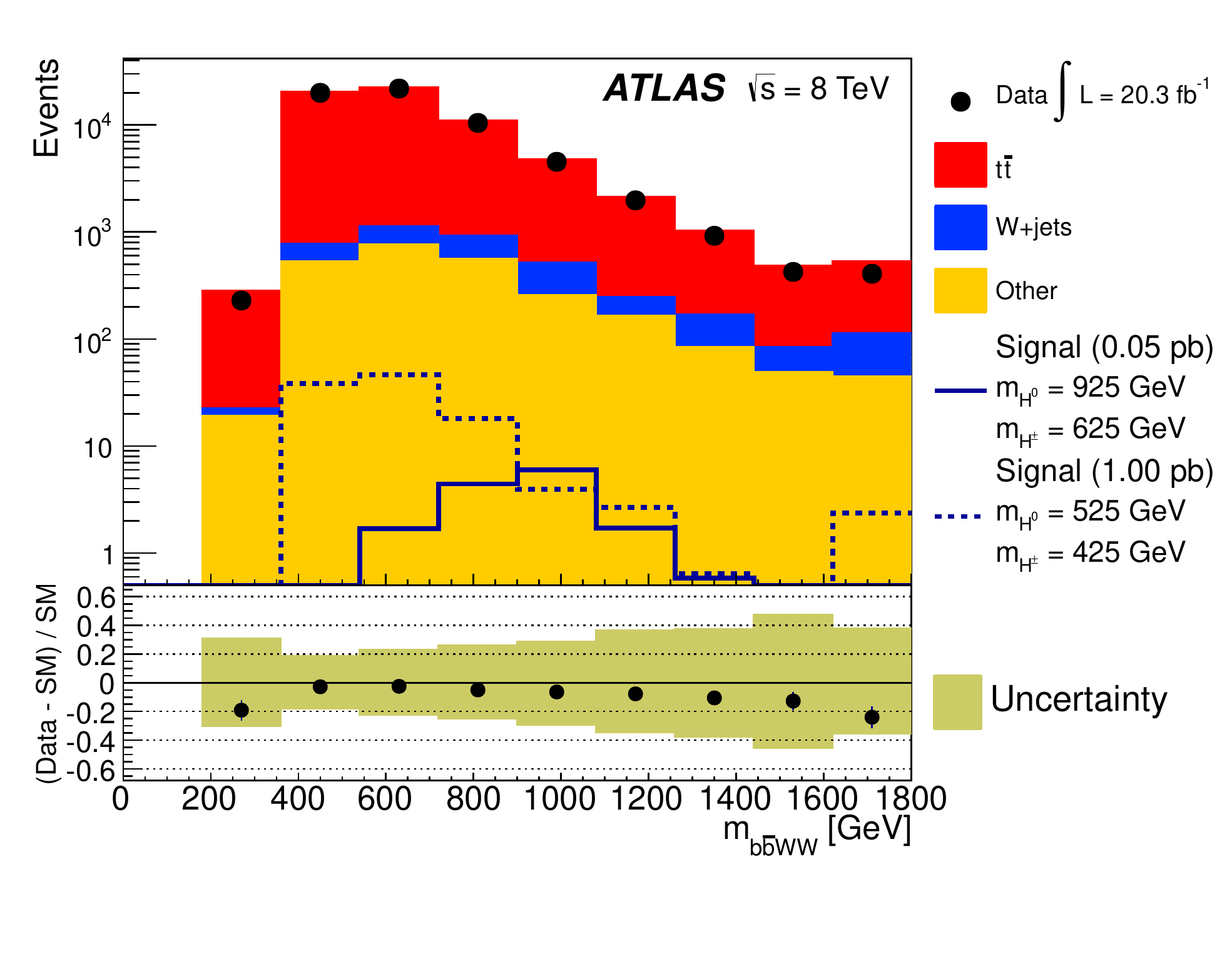}
\caption{Distributions of $m_{b\bar{b}}$ (top) and $m_{b\bar{b}WW}$ (bottom) with uncertainties in the
  control region CR3. The data (black points) are compared to the
  background model (stacked histogram). The final bin contains any
  overflow events. Two choices of signal hypotheses are also shown. }
\label{fig:control_cr3}
\end{center}
\end{figure}

Figs.~\ref{fig:control_cr2} and ~\ref{fig:control_cr3} illustrate the modelling of the Higgs-boson
mass reconstruction in CR1, CR2 and CR3. The data and simulation agree within total
uncertainties over the entire phase space. This is important, as the BDT may
utilize any part of this phase space to build a powerful discriminant. In addition, the BDT output in each of the three control regions is compared to the predicted output and found to agree within statistical and systematic uncertainties.

\renewcommand{\arraystretch}{1.0} 
 \begin{table} 
 \begin{center} 
 \begin{tabular}{lcc} 
 \hline \hline 
 Uncertainty & Background & Signal \\  
 & Yields (\%) & Efficiency (\%)\\
 & & {\scriptsize $m_{H^0} = 425$ GeV }  \\  
 &  & {\scriptsize $m_{H^\pm} = 225$ GeV }  \\  \hline 
Jet vertex fraction & $\pm$ 1.6  & $\pm$ 2.1  \\ 
$b$-tagging  eff. & $\pm$ 8.8  & $\pm$ 14  \\ 
Jet energy scale & $\pm$ 3.9  & $\pm$ 7  \\ 
Jet energy res. & $\pm$ 1.1  & $\pm$ 11  \\ 
Jet reco. eff. & $\pm$ $<$1.0  & $\pm$ $<$1.0  \\ 
$\mu$ momentum  & $\pm$ $<$1.0  & $\pm$ $<$1.0  \\ 
$e$ energy  & $\pm$ $<$1.0  & $\pm$ $<$1.0  \\ 
Lepton trigger eff.  & $\pm$ $<$1.0  & $\pm$ 1.8  \\ 
Lepton ident. eff.  & $\pm$ 1.5  & $\pm$ 2.1  \\ 
Lepton reco. eff. & $\pm$ $<$1.0  & $\pm$ $<$1.0  \\ 
$W$-boson+jets shape  & $\pm$ $<$1.0 & - \\ 
Quark/gluon radiation & $\pm$ $<$1.0  & $\pm$ 2.8  \\ 
\ttbar modelling & $\pm$ 2.7 & - \\ 
Bg. normalization & $\pm$ 5.5  & -  \\ 
Luminosity & $\pm$ 2.8  & $\pm$ 2.8  \\ 
 \hline 
Total uncertainty & $\pm$ 12  & $\pm$ 20  \\ 
\hline \hline 
 \end{tabular} 
\caption{Details  of the systematic uncertainties relative to the total expected 
background  and the signal efficiency in the signal region for a
Higgs-boson cascade signal 
with $m_{H^0}, m_{H^{\pm}}$ = 425, 225 GeV. The signal region for this mass point is defined 
as the events that pass the BDT threshold  for this mass sample. The positive and negative 
relative shifts have been averaged for compactness.}
\label{tab:syst} 
 \end{center} 
  \end{table} 
\renewcommand{\arraystretch}{1.4}

\section{Systematic uncertainties}
\label{sec:syst}
Several sources of systematic uncertainties are relevant to this
analysis.

Instrumental systematic 
uncertainties are related to the reconstruction of physics
objects. For jets, systematic uncertainties on the jet energy scale,
energy resolution, and reconstruction efficiency  are
included. For leptons, the systematic uncertainties from the momentum or energy scale and resolution, trigger efficiency,
reconstruction, and  identification efficiency are incorporated.
Systematic uncertainties related to the performances of the $b$-tagging and JVF requirements are also included. 

Due to the presence of multiple ($\geq$4)
jets and the dominant \ttbar background  (roughly 90\% in the signal
region), significant systematic uncertainties are associated with jets and the modelling
of the \ttbar background. Table~\ref{tab:syst} lists the impact of these uncertainties on the background estimates and signal efficiency for an example signal region given by the BDT threshold for a signal with  $m_{H^0}, m_{H^\pm} = 425, 225$ GeV.

Several sources of uncertainty on the jet energy scale calibration are considered, 
such as uncertainties due to pile-up and the light-quark and gluon composition. These sources are added in quadrature and listed 
as one systematic uncertainty in Table~\ref{tab:syst}. As a further uncertainty, the 
jet energy is smeared to cover any disagreements in the jet energy resolution measured
in data and in simulated event samples. A jet reconstruction efficiency~\cite{jer} systematic uncertainty is applied
by randomly discarding a fraction of low-$\pT$ jets in the simulated  events. The jet $b$-tagging efficiencies are evaluated in data
and MC~\cite{btag}. The difference is corrected with a scale factor, the uncertainty of which is treated as a
systematic uncertainty. A small discrepancy in the efficiency
of the JVF requirement has been observed between data and MC
simulation. The JVF requirement is varied to cover this observed
discrepancy, and the resulting change in the expected background is taken as a systematic uncertainty.

Systematic uncertainties associated with leptons are found to have a small effect, typically less than 
1\% relative to background estimates and signal efficiency. For muons, the uncertainty in the momentum scale and resolution is accounted for. For electrons,
the uncertainties in the energy scale and resolution are included. For both leptons, uncertainties on
the trigger, identification, and reconstruction efficiencies are incorporated.

The uncertainty due to the modelling of initial- and final-state quark
and gluon radiation (ISR/FSR)
is estimated using \ttbar\ events produced with the {\sc AcerMC} generator interfaced 
with {\sc PYTHIA}, where the parameters controlling ISR/FSR are varied
in a range suggested by the data in  the analysis of Ref.~\cite{ rapidity_gap1}. For the signal, events generated with varied
ISR/FSR parameters in {\sc PYTHIA} are compared to the nominal
simulation;  the differences in 
background yields and signal efficiency estimates are taken as a systematic uncertainty.

The systematic uncertainty due to the modelling of  \ttbar\ production  is
estimated by comparing results obtained with {\sc MC@NLO}, {\sc POWHEG},
and {\sc ALPGEN} signal samples. An uncertainty due to the 
theoretical $t\bar{t}$ cross section~\cite{ttbar_xsec} is applied to the 
overall $t\bar{t}$ normalization. Since this is the dominant background the effect on 
the total background uncertainty is substantial (about 5\% relative to
the background estimate); the total normalization uncertainty on the
background is 5.5\%.

Since non-\ttbar processes account for less than 10\% of the background
in the signal region, systematic uncertainties associated with them are found to have a small impact on the overall background uncertainty.
The systematic uncertainty related to the modelling of $W$-boson+jets is determined by
varying the parameterization of the renormalization and factorization scales in {\sc ALPGEN}. As default, both scales are set to $(m_W^2+(p_{{\textrm T}}^W)^2)$ and this is varied by factors of two and by changing the form to $(m_W^2+\sum_{\textrm{jets}} p_{\textrm T}^2)$.
This systematic uncertainty is found to be small ($<$1\%). 
An overall uncertainty of 4\% is applied to the $W$-boson + jets estimate due to uncertainties in the the cross section, with an additional
24\% per jet added in quadrature due to the uncertainty in Berends scaling~\cite{berends}. This results in a 48\% 
uncertainty for events with four jets, contributing to the overal 5.5\% uncertainty on the background normalization.

The systematic uncertainty 
due to single top-quark, diboson, and $Z$-boson+jets production is
evaluated by varying their cross sections within their uncertainties as described in Ref.~\cite{Aad:2010ey}.
Since these contributions are small, the systematics associated with them are found to be negligible ($<$1\%).

Finally, the luminosity uncertainty, measured using techniques similar to those 
described in Ref.~\cite{Luminosity:2013}, is 2.8\%.

\begin{figure}[!h]
\begin{center}
\includegraphics[width=0.85\columnwidth]{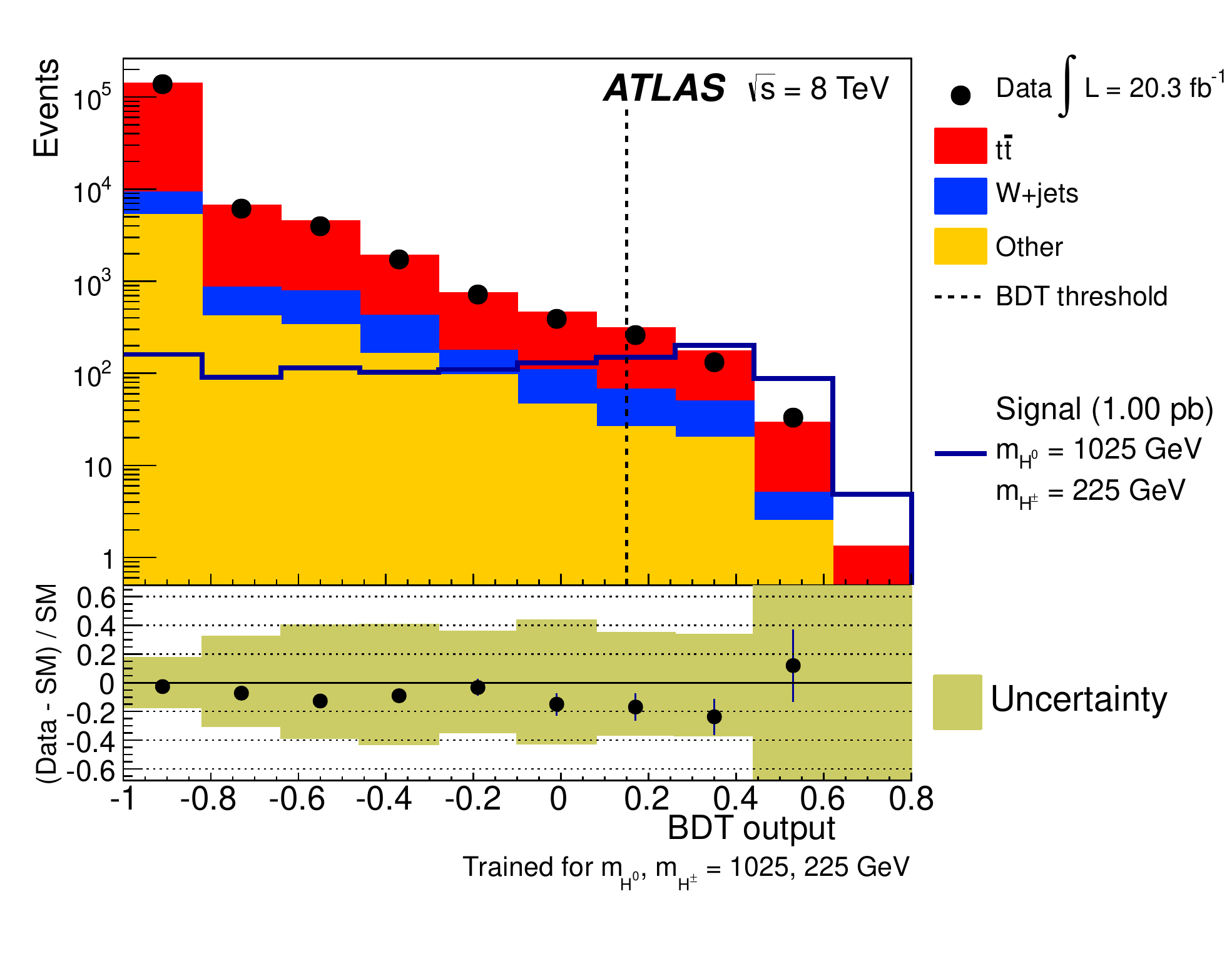}\\
\includegraphics[width=0.85\columnwidth]{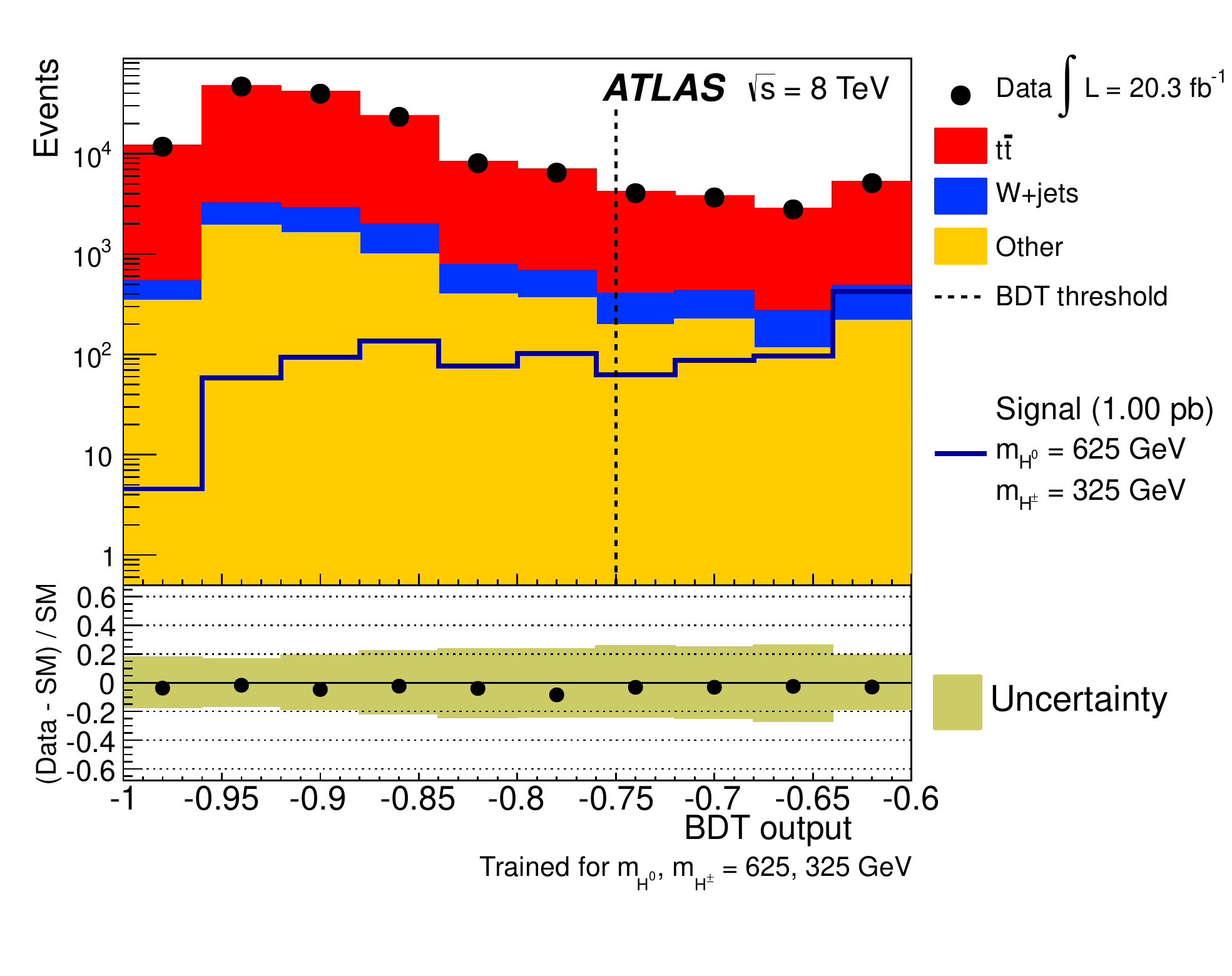}\\
\includegraphics[width=0.85\columnwidth]{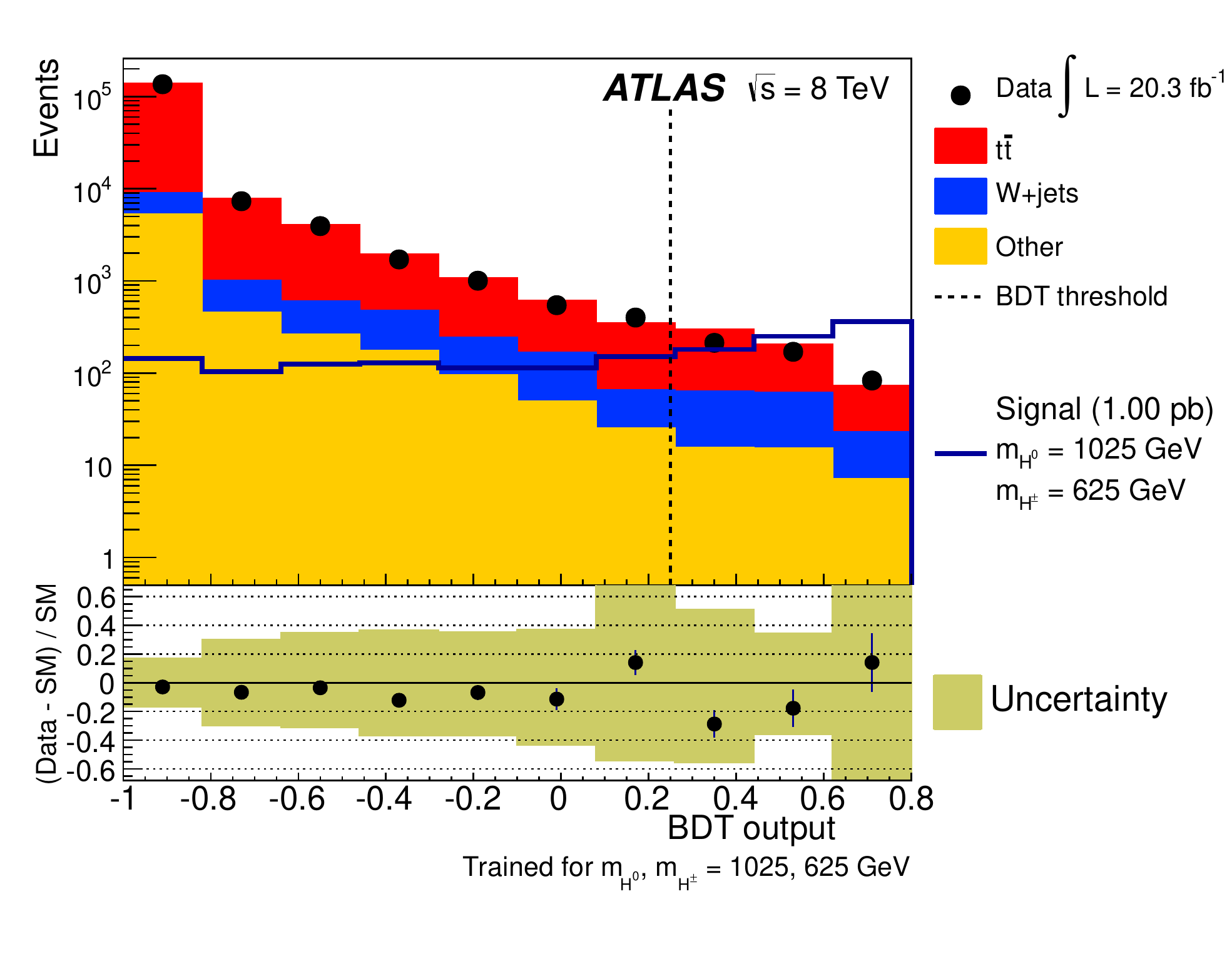}
\caption{Distribituions of the BDT output in the signal regions for three example signal mass points, 
$m_{H^0}$, $m_{H^\pm}$ = $1025$, $225$ GeV (top), $m_{H^0}$, $m_{H^\pm}$ = $625$, $325$ GeV (middle), 
$m_{H^0}$, $m_{H^\pm}$ = $1025$, $625$ GeV (bottom). Signal histograms
have been scaled to a production cross section of 1 pb. BDT
thresholds are  shown as dashed lines for each mass point. The background model is shown
  as the coloured stacked histogram. The final bin contains any overflow events.}
\label{fig:SR_disc}
\end{center}
\end{figure}

\begin{figure}[!h]
\begin{center}
\includegraphics[width=0.9\columnwidth]{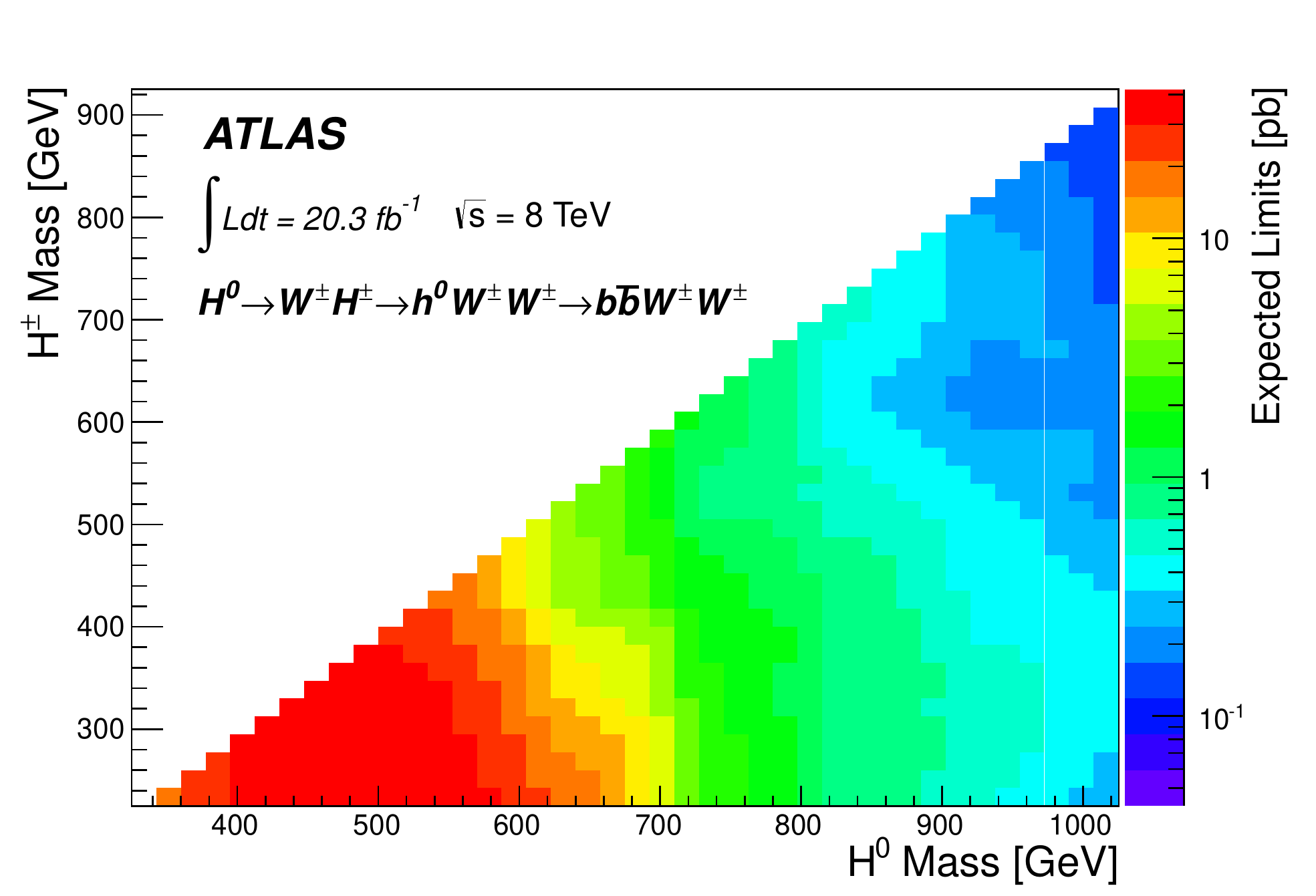}\\
\includegraphics[width=0.9\columnwidth]{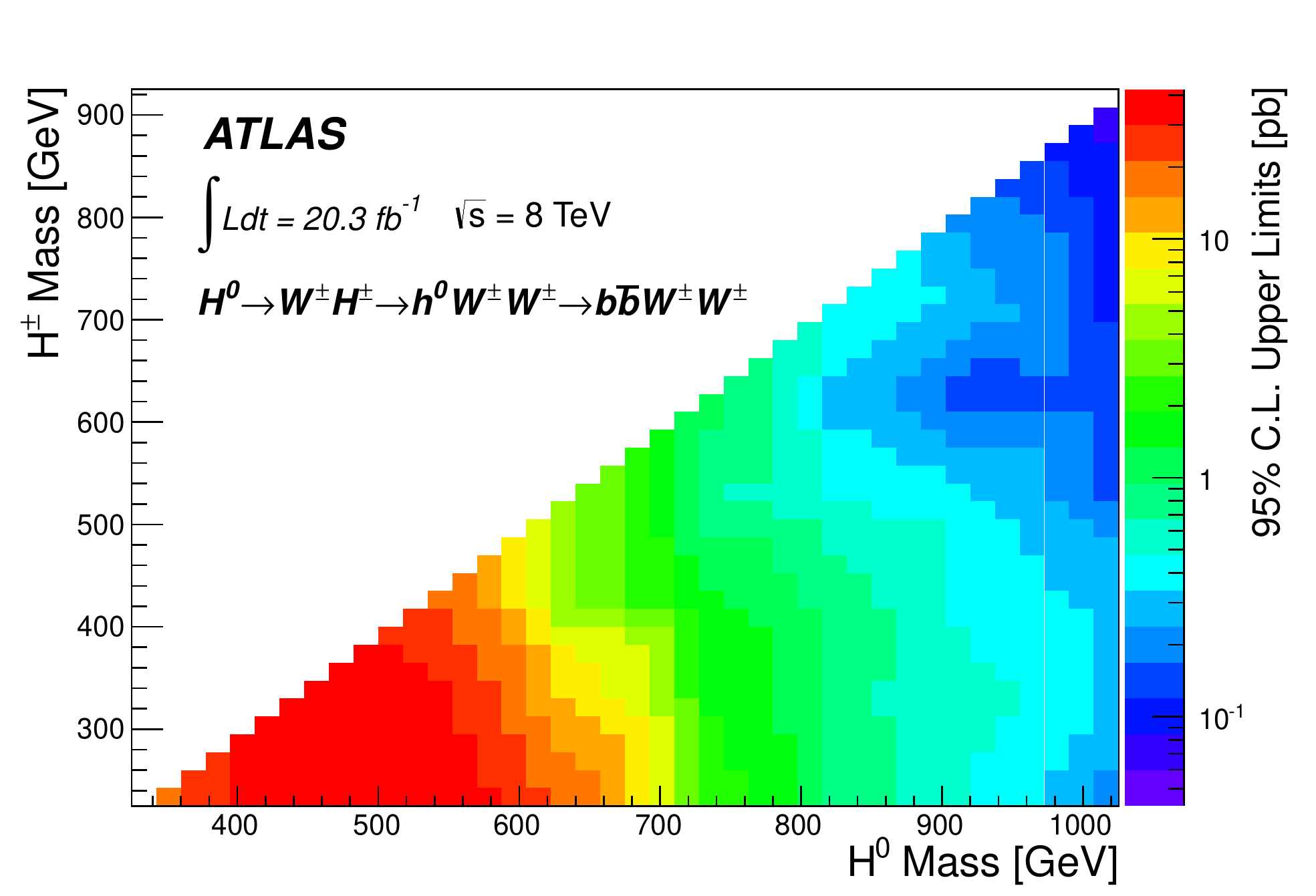}\\
\includegraphics[width=0.9\columnwidth]{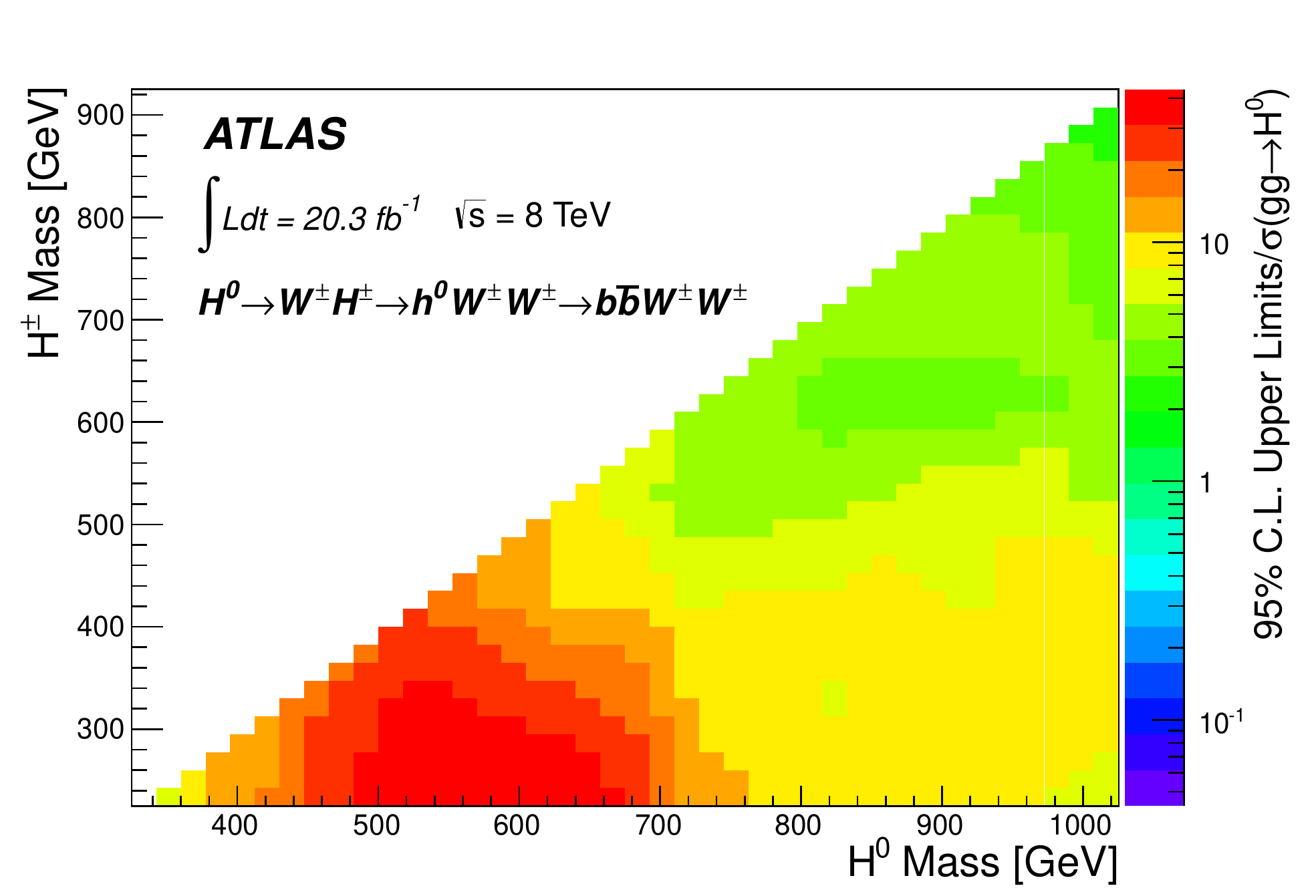}
\caption{The expected (top) and observed (middle) 95\% C.L. upper
  limits on the cross section for $gg \to H^0 \to
  W^\mp H^\pm \to W^\pm W^\mp h^0 \to W^\pm  W^\mp b\bar{b}$ as a function of  $m_{H^0}$ and $m_{H^\pm}$.
  The ratio  (bottom) of the observed 95\% C.L. upper limits on the cross
  section to the theoretical cross section for a heavy Higgs boson 
produced via gluon-gluon fusion at the SM rate.}
\label{fig:res}
\end{center}
\end{figure}

\section{Results}
\label{sec:results}
The yields in the signal regions are given in Table~\ref{tab:summary}. The 
observed yields are found to be consistent with SM background
expectations, within uncertainties. The BDT outputs for three example
signal mass points are illustrated in Fig.~\ref{fig:SR_disc}.

\begin{table*}[ht]
\caption{Expected background and observed yield, with expected and
  observed cross-section upper limits for each signal hypothesis in their
  respective signal regions. For the expected cross-section limits,
  the uncertainties describe a range which contains the limits in 68\%
  and 95\% of simulated experiments, respectively.
Also shown is the $p$-value for the background-only hypothesis.}
\label{tab:summary}
\begin{center}
\begin{tabular}{ll cccc c rrr}
\hline
\hline
\multicolumn{2}{c}{Masses [GeV]} &\multicolumn{3}{c}{Yields}&
\multicolumn{1}{c}{Efficiency (\%)}&
\multicolumn{2}{c}{Limits [pb] at 95\% C.L.}& Backgr.-only\\
 $m_{H^0}$ & $m_{h^\pm}$ &$t\bar{t}$ &Total Bkgd. & Obs.& Signal &Expected &
 Obs.& $p$-value\\
\hline
325 & 225 &  $8.4^{+1.0}_{-1.0}$ $\cdot 10^{3}$ &  $ 8.9^{+1.0}_{-1.3}$ $\cdot 10^{3}$ & 9244& $0.96^{+0.32}_{-0.32}$ & $11^{+5 +12}_{-3 -5}$ &12.8 & 0.36 \\
425 & 225 &  $9.8^{+1.2}_{-1.2}$  $\cdot 10^{4}$ &  $ 10.3^{+1.2}_{-1.5}$ $\cdot 10^{4}$& 103738& $2.9^{+0.8}_{-0.8}$ & $41^{+18 +40}_{-12 -19}$ &40.2 & 0.49 \\
425 & 325 &  $9.4^{+1.1}_{-1.1}$ $\cdot 10^{4}$ &  $ 9.8^{+1.1}_{-1.4}$ $\cdot 10^{4}$ & 99770& $2.9^{+0.7}_{-0.7}$ & $40^{+17 +40}_{-11 -18}$ &39 & 0.49 \\
525 & 225 &  $11.4^{+1.6}_{-1.5}$$\cdot 10^{4}$ &  $ 12.3^{+1.6}_{-1.9}$ $\cdot 10^{4}$& 124802& $3.8^{+0.8}_{-0.8}$ & $42^{+17 +41}_{-12 -19}$ &42.8 & 0.49 \\
525 & 325 &  $11.5^{+1.6}_{-1.6}$ $\cdot 10^{4}$&  $ 12.6^{+1.7}_{-1.9}$ $\cdot 10^{4}$& 127702& $4.6^{+1.0}_{-1.0}$ & $35^{+15 +34}_{-10 -16}$ &51.7 & 0.49 \\
525 & 425 &  $41.0^{+6.3}_{-6.0}$  $\cdot 10^{2}$ &  $ 44.0^{+6.9}_{-7.1}$ $\cdot 10^{2}$& 4342& $0.31^{+0.1}_{-0.1}$ & $23^{+11 +28}_{-7 -11}$ &22.6 & 0.49 \\
625 & 225 &  $20.3^{+2.9}_{-3.0}$   $\cdot 10^{3}$& $ 23.0^{+3.4}_{-4.0}$  $\cdot 10^{3}$& 22907& $1.6^{+0.4}_{-0.4}$ & $21^{+8 +20}_{-6 -9}$ &19.6 & 0.5 \\
625 & 325 &  $10.0^{+1.5}_{-1.3}$ $\cdot 10^{3}$ &  $ 10.8^{+1.7}_{-1.7}$ $\cdot 10^{3}$& 11064& $1.5^{+0.3}_{-0.4}$ & $11^{+5 +11}_{-3 -5}$ &11 & 0.49 \\
625 & 425 &  $20.5^{+3.4}_{-3.0}$ $\cdot 10^{2}$ &  $ 24.4^{+4.1}_{-4.9}$$\cdot 10^{2}$  & 2294& $0.85^{+0.27}_{-0.22}$ & $4.8^{+2.1 +5.2}_{-1.4 -2.2}$ &4.27 & 0.5 \\
625 & 525 &  $22.0^{+3.4}_{-3.7}$ $\cdot 10^{2}$&  $ 25.3^{+4.4}_{-5.0}$ $\cdot 10^{2}$& 2564& $1.0^{+0.3}_{-0.2}$ & $4.3^{+1.8 +4.5}_{-1.2 -2.0}$ &4.22 & 0.5 \\
725 & 225 &  $31.8^{+5.2}_{-5.2}$ $\cdot 10^{2}$&  $ 37.7^{+6.9}_{-7.6}$ $\cdot 10^{2}$& 3710& $2.4^{+0.6}_{-0.6}$ & $2.6^{+1.1 +2.6}_{-0.7 -1.2}$ &2.42 & 0.5 \\
725 & 325 &  $36.0^{+5.2}_{-5.7}$ $\cdot 10^{2}$&  $ 41.0^{+7.0}_{-7.8}$ $\cdot 10^{2}$& 3980& $2.7^{+0.6}_{-0.6}$ & $2.1^{+0.9 +2.2}_{-0.6 -1.0}$ &1.99 & 0.5 \\
725 & 425 &  $24.9^{+4.3}_{-3.9}$ $\cdot 10^{2}$&  $ 29.6^{+5.7}_{-6.2}$ $\cdot 10^{2}$& 2828& $2.8^{+0.7}_{-0.6}$ & $1.7^{+0.7 +1.8}_{-0.5 -0.8}$ &1.49 & 0.5 \\
725 & 525 &  $13.4^{+2.1}_{-1.9}$ $\cdot 10^{2}$&  $ 16.3^{+3.3}_{-3.5}$$\cdot 10^{2}$ & 1538& $2.8^{+0.7}_{-0.6}$ & $0.84^{+0.40 +1.00}_{-0.25 -0.40}$ &0.718 & 0.5 \\
725 & 625 &  $23.6^{+3.6}_{-3.5}$ $\cdot 10^{2}$&  $ 28.7^{+5.3}_{-6.1}$ $\cdot 10^{2}$& 2702& $3.5^{+0.9}_{-0.9}$ & $1.2^{+0.6 +1.4}_{-0.4 -0.6}$ &1.06 & 0.5 \\
825 & 225 &  $7.1^{+0.92}_{-1.3}$  $\cdot 10^{2}$&  $ 8.9^{+1.4}_{-2.4}$  $\cdot 10^{2}$& 830& $1.4^{+0.4}_{-0.4}$ & $0.91^{+0.42 +1.00}_{-0.27 -0.43}$ &0.794 & 0.5 \\
825 & 325 &  $10.8^{+1.5}_{-1.7}$  $\cdot 10^{2}$&  $ 13.0^{+2.4}_{-2.8}$  $\cdot 10^{2}$& 1237& $2.4^{+0.6}_{-0.6}$ & $0.82^{+0.36 +0.89}_{-0.23 -0.38}$ &0.717 & 0.5 \\
825 & 425 &  $10.5^{+1.9}_{-1.6}$  $\cdot 10^{2}$&  $ 12.7^{+2.7}_{-2.5}$  $\cdot 10^{2}$& 1186& $2.3^{+0.6}_{-0.5}$ & $0.92^{+0.41 +1.00}_{-0.26 -0.42}$ &0.8 & 0.5 \\
825 & 525 &  $8.0^{+1.6}_{-1.3}$ $\cdot 10^{2}$&  $ 9.9^{+2.4}_{-2.8}$ $\cdot 10^{2}$& 901& $2.9^{+0.7}_{-0.7}$ & $0.55^{+0.25 +0.62}_{-0.16 -0.26}$ &0.457 & 0.5 \\
825 & 625 &  $5.9^{+0.8}_{-1.0}$ $\cdot 10^{2}$  &$ 7.7^{+1.6}_{-1.8}$ $\cdot 10^{2}$& 696& $2.5^{+0.7}_{-0.7}$ & $0.36^{+0.18 +0.45}_{-0.11 -0.18}$ &0.282 & 0.5 \\
825 & 725 &  $5.1^{+0.8}_{-0.7}$ $\cdot 10^{2}$&  $ 6.6^{+1.6}_{-1.3}$ $\cdot 10^{2}$& 628& $1.6^{+0.5}_{-0.4}$ & $0.56^{+0.28 +0.71}_{-0.17 -0.27}$ &0.484 & 0.5 \\
925 & 225 &  $5.7^{+0.9}_{-0.8}$ $\cdot 10^{2}$&  $ 7.0^{+1.3}_{-1.4}$ $\cdot 10^{2}$& 641& $2.1^{+0.5}_{-0.5}$ & $0.51^{+0.22 +0.55}_{-0.14 -0.24}$ &0.441 & 0.5 \\
925 & 325 &  $7.4^{+1.1}_{-1.2}$ $\cdot 10^{2}$&  $ 9.3^{+1.4}_{-1.9}$ $\cdot 10^{2}$& 876& $2.7^{+0.7}_{-0.6}$ & $0.58^{+0.27 +0.69}_{-0.17 -0.27}$ &0.528 & 0.5 \\
925 & 425 &  $6.6^{+1.0}_{-1.0}$ $\cdot 10^{2}$&  $ 8.2^{+1.6}_{-1.7}$ $\cdot 10^{2}$& 796& $3.1^{+0.7}_{-0.7}$ & $0.40^{+0.18 +0.46}_{-0.12 -0.19}$ &0.38 & 0.5 \\
925 & 525 &  $6.0^{+1.0}_{-1.1}$ $\cdot 10^{2}$&  $ 8.1^{+1.8}_{-1.9}$ $\cdot 10^{2}$& 787& $3.3^{+0.9}_{-0.8}$ & $0.37^{+0.17 +0.43}_{-0.11 -0.18}$ &0.345 & 0.5 \\
925 & 625 &  $1.8^{+0.4}_{-0.3}$ $\cdot 10^{2}$&  $ 2.4^{+0.6}_{-0.6}$ $\cdot 10^{2}$& 185& $2.4^{+0.6}_{-0.6}$ & $0.17^{+0.08 +0.20}_{-0.05 -0.08}$ &0.123 & 0.5 \\
925 & 725 &  $2.8^{+0.7}_{-0.4}$ $\cdot 10^{2}$&  $ 3.8^{+1.0}_{-0.9}$ $\cdot 10^{2}$& 359& $2.7^{+0.7}_{-0.7}$ & $0.30^{+0.14 +0.34}_{-0.09 -0.14}$ &0.272 & 0.5 \\
925 & 825 &  $4.7^{+0.7}_{-0.7}$ $\cdot 10^{2}$&  $ 6.0^{+1.3}_{-1.2}$ $\cdot 10^{2}$& 537& $3.6^{+1.0}_{-0.9}$ & $0.22^{+0.11 +0.28}_{-0.07 -0.11}$ &0.172 & 0.5 \\
1025 & 225 &  $2.9^{+0.5}_{-0.7}$ $\cdot 10^{2}$&  $ 3.7^{+1.1}_{-1.2}$ $\cdot 10^{2}$& 306& $1.9^{+0.5}_{-0.5}$ & $0.23^{+0.11 +0.28}_{-0.07 -0.11}$ &0.15 & 0.5 \\
1025 & 325 &  $7.1^{+1.0}_{-1.2}$ $\cdot 10^{2}$&  $ 9.4^{+2.0}_{-2.2}$ $\cdot 10^{2}$& 839& $3.3^{+0.8}_{-0.8}$ & $0.37^{+0.17 +0.41}_{-0.11 -0.17}$ &0.292 & 0.5 \\
1025 & 425 &  $5.4^{+0.8}_{-0.9}$ $\cdot 10^{2}$&  $ 7.1^{+1.5}_{-1.7}$$\cdot 10^{2}$& 691& $3.3^{+0.8}_{-0.8}$ & $0.33^{+0.15 +0.37}_{-0.09 -0.15}$ &0.308 & 0.5 \\
1025 & 525 &  $2.4^{+0.5}_{-0.6}$ $\cdot 10^{2}$&  $ 3.5^{+1.6}_{-1.1}$$\cdot 10^{2}$ & 297& $3^{+0.8}_{-0.7}$ & $0.19^{+0.09 +0.22}_{-0.05 -0.09}$ &0.147 & 0.5 \\
1025 & 625 &  $4.3^{+0.7}_{-0.9}$ $\cdot 10^{2}$&  $ 5.7^{+1.6}_{-1.4}$ $\cdot 10^{2}$& 477& $3.6^{+1.0}_{-0.9}$ & $0.19^{+0.10 +0.25}_{-0.06 -0.09}$ &0.132 & 0.5 \\
1025 & 725 &  $1.4^{+0.2}_{-0.3}$ $\cdot 10^{2}$&  $ 2.0^{+0.5}_{-0.6}$$\cdot 10^{2}$ & 162& $1.8^{+0.5}_{-0.4}$ & $0.15^{+0.08 +0.19}_{-0.05 -0.07}$ &0.0939 & 0.5 \\
1025 & 825 &  $2.1^{+0.3}_{-0.5}$ $\cdot 10^{2}$&  $ 3.0^{+0.7}_{-1.}$ $\cdot 10^{2}$& 241& $2.6^{+0.6}_{-0.6}$ & $0.14^{+0.07 +0.18}_{-0.04 -0.07}$ &0.0887 & 0.5 \\
1025 & 925 &  $9.4^{+1.2}_{-1.7}$ $\cdot 10$&  $ 13.7^{+4.7}_{-4.4}$ $\cdot 10$& 110& $1.9^{+0.5}_{-0.5}$ & $0.10^{+0.06 +0.15}_{-0.03 -0.05}$ &0.0653 & 0.5 \\
\hline
\hline
\end{tabular}
\end{center}
\end{table*}

 The 95\% confidence-level production cross-section upper limits for the various signal 
hypotheses are obtained using the CLs frequentist method~\cite{cls},
with the profile likelihood ratio of the number of events that pass
the BDT threshold as the test statistic~\cite{asymp} as implemented in Ref.~\cite{roostats}. Systematic 
uncertainties are treated as nuisance parameters and the calculation uses the
asymptotic approximation~\cite{asymp}. Table~\ref{tab:summary} presents the signal
efficiencies, the total expected background and observed event counts for each signal case, as 
well as the expected and observed limits with the local $p$-values. The $p$-values are defined as the probabilities
under the background-only hypothesis to observe these data or  data which are more signal-like. The $p$-values have a maximum 
possible value of $0.5$, which is the case when $n<b$, where $b$ is the number of events expected from the background model and $n$ is the number
of events observed in the data.

Since the signal regions are correlated, background-only
pseudo-experiments are used to estimate the expected distribution of
the $p$-values in all the signal regions, accounting for the correlations. The observed distribution of $p$-values is found to be
consistent with the expectation from pseudo-experiments. The expected and observed limits as a function of the $H^0$ and $H^\pm$ masses are illustrated in Fig.~\ref{fig:res}. 
The limits are the weakest in low Higgs-boson mass regions due to the poorer separation between $t\bar{t}$ and 
signal events.

In order to facilitate the comparison of these results with those obtained by other experiments, the observed cross-section limits are compared to
the predictions for a heavy Higgs boson with SM-like $gg$-fusion production (Fig.~\ref{fig:res}). The theoretical production cross section
of a heavy SM-like Higgs boson 
(only gluon fusion is considered) is calculated 
in the complex-pole scheme using the dFG~\cite{higgs_xsec} program, to next-to-next-to-leading order (NNLO) in QCD. Next-to-leading order (NLO) electroweak (EW) corrections are also applied, as well as QCD soft-gluon resummations up to next-to-next-to-leading log (NNLL).Using this benchmark, the cross-section upper limits observed are
greater than the theoretical cross sections of the 
heavy Higgs boson, $H^{0}$, for all mass points tested. Therefore, the current limits are not stringent enough to 
exclude models with SM-like production rates even with 100\% branching ratios for both 
$H^{0} \rightarrow H^{\pm}W^{\pm}$ and $H^{\pm} \rightarrow
h^{0}W^{\pm}$ and SM values for BR($h^0\rightarrow b\bar{b}$). The limits are most stringent in the high $H^{0}$  and $H^{\pm}$ mass regions, 
where the ratio of the limits to the theoretical cross section is nearly unity. 
This search produces tighter bounds than those obtained by the CDF
collaboration~\cite{cdfprl}. 

Additionally,  the results of this search are interpretted
in the context of a heavy CP-even Higgs boson
of a type-II 2-Higgs-Doublet Model~\cite{Gunion:1989we}
produced via gluon fusion.
This model has seven free parameters: the mass of the
CP-even Higgs bosons ($m_{h^0}$ and $m_{H^0}$), the mass of the CP-odd
Higgs boson ($m_{A}$), the mass of the charged scalar ($m_{H^{\pm}}$),
the mixing angle between the CP-even Higgs bosons ($\alpha$),
the ratio of the vacuum expectation values of the two Higgs
doublets ($\tan\beta$), and the $Z_{2}$-symmetry soft-breaking-term
coefficient of the Higgs potential ($\mathcal{M}_{12}^{2}$).
The parameter space of the type-II 2HDM is sampled
for given values of $m_{H^0}$ and $m_{H^{\pm}}$
and assuming  $m_{h^0}=125$ GeV and $\sin(\beta - \alpha) \ge 0.99$.
The latter assumptions are made in order to maintain a SM-like
Higgs boson with properties similar to those observed at the
LHC. The gluon-fusion production cross section is calculated
with SusHi~\cite{sushi} at NNLO precision in QCD corrections,
and the branching ratio of the cascade $H^0 \rightarrow W^\mp H^\pm
\rightarrow W^+W^-h \rightarrow W^+W^-b\bar{b}$ with 2HDMC \cite{2HDMC}.
Only parameter space points that satisfy theory constraints
are considered. The theory constraints include Higgs-potential stability, tree-level
unitarity for Higgs-boson scattering~\cite{Ginzburg:2005dt},
and the perturbative nature of the quartic Higgs-boson couplings,
as these are implemented in 2HDMC.
The type-II 2HDM phase space is scanned with a million
random points per $(m_{H^0},m_{H^{\pm}})$ pair.
The majority of the spanned phase space violates the
theoretical constraints mentioned above. The points
with the lowest cross section times branching fraction 
$\sigma \times$BF(excluded)$/ \sigma \times$BF(theory) which satisfy the above constraints
are shown in Table~\ref{tab:2hdm}, where $\sigma$
is the cross section and BF is the branching fraction.  None are excluded by the
limits presented here.

\begin{table*}[ht]
\caption{Interpretation of the results in some type-II
2HDM parameter space choices. For each value of $m_{H^0},m_{H^\pm}$, where at
least one valid point is found, sample points in the space of the
parameters ($\tan(\beta)$,  $\sin(\beta-\alpha)$, $m_A$, and
  $\mathcal{M}_{12}^2$) which satisfy potential stability, unitarity and
perturbativity constraints and give the smallest ratio of excluded to
predicted cross section are shown.}
\label{tab:2hdm}
\begin{center}
\begin{tabular}{llllllrrr}
\hline\hline
$m_{H^0}$ & $m_{H^\pm}$ & $\tan(\beta)$ & $\sin(\beta-\alpha)$ & $m_A$
& $\mathcal{M}_{12}^2$ &
$\sigma(H^0)$ & BF($H^{0} \rightarrow h^0W^+W^-$) & Excl/Pred \\
{[GeV]} & [GeV] & & & [GeV] &[TeV$^{2}$] & [pb] \\
\hline
325     & 225    & 15    & 0.99          & 303   & 6.9  $\cdot 10^{-3}$        & 28    & 0.222         & 2.1\\
425     & 225    & 20    & 0.99          & 439   & 8.9 $\cdot 10^{-3}$         & 2     & 0.404         & 41\\
425     & 325    & 10    & 0.99          & 486   & 1.8  $\cdot 10^{-2}$        & 10    & 0.288         & 14\\
525     & 325    & 10    & 0.99          & 384   & 2.7 $\cdot 10^{-2}$         & 3     & 0.436         & 39\\
525     & 425    & 10    & 0.99          & 384   & 2.7 $\cdot 10^{-2}$         & 5     & 0.136         & 34\\
625     & 325    & 10    & 0.99          & 549   & 3.9 $\cdot 10^{-2}$         & 1     & 0.501         & 20\\
625     & 425    & 10    & 0.99          & 693   & 3.9 $\cdot 10^{-2}$         & 2     & 0.607         & 4.1\\
625     & 525    & 10    & 0.99          & 693   & 3.9 $\cdot 10^{-2}$         & 3     & 0.219         & 7.7\\
725     & 325    & 1     & 0.99          & 675   & 5.9 $\cdot 10^{-2}$         & 0.3& 0.009         & 664\\
725     & 425    & 10    & 0.99          & 731   & 5.2 $\cdot 10^{-2}$         & 1     & 0.643         & 3.5\\
725     & 525    & 10    & 0.99          & 731   & 5.2 $\cdot 10^{-2}$         & 1     & 0.659         & 1.1\\
725     & 625    & 10    & 0.99          & 396   & 5.2 $\cdot 10^{-2}$         & 1& 0.002         & 440\\
825     & 525    & 1     & 0.99          & 788   & 1.3 $\cdot 10^{-1}$        & 0.3     & 0.024         & 76\\
825     & 625    & 1     & 0.99          & 788   & 1.3 $\cdot 10^{-1}$        & 0.3     & 0.021         & 41\\
825     & 725    & 10    & 0.999         & 807   & 6.8  $\cdot 10^{-2}$        & 1     & 0.168         & 4.1\\
925     & 725    & 1     & 0.999         & 921   & 2.4   $\cdot 10^{-1}$      & 0.2     & 0.003         & 530\\
1025    & 825    & 1     & 0.999         & 920   & 3.4  $\cdot 10^{-1}$       & 0.1     & 0.003         & 243\\
\hline\hline
\end{tabular}
\end{center}
\end{table*}

In conclusion, the first LHC search for a topology in which a heavy Higgs boson decays via a cascade of lighter charged and neutral Higgs bosons has been performed by the ATLAS experiment using data corresponding to an integrated luminosity of 20.3 fb$^{-1}$ in $pp$ collisions at $\sqrt{s} = 8$~TeV. 
No significant excess of events above the expectation from the SM background was found and limits on the 
production cross section have been set. 

\clearpage

\section{Acknowledgements}
We thank CERN for the very successful operation of the LHC, as well as the
support staff from our institutions without whom ATLAS could not be
operated efficiently.

We acknowledge the support of ANPCyT, Argentina; YerPhI, Armenia; ARC,
Australia; BMWF and FWF, Austria; ANAS, Azerbaijan; SSTC, Belarus; CNPq and FAPESP,
Brazil; NSERC, NRC and CFI, Canada; CERN; CONICYT, Chile; CAS, MOST and NSFC,
China; COLCIENCIAS, Colombia; MSMT CR, MPO CR and VSC CR, Czech Republic;
DNRF, DNSRC and Lundbeck Foundation, Denmark; EPLANET, ERC and NSRF, European Union;
IN2P3-CNRS, CEA-DSM/IRFU, France; GNSF, Georgia; BMBF, DFG, HGF, MPG and AvH
Foundation, Germany; GSRT and NSRF, Greece; ISF, MINERVA, GIF, DIP and Benoziyo Center,
Israel; INFN, Italy; MEXT and JSPS, Japan; CNRST, Morocco; FOM and NWO,
Netherlands; BRF and RCN, Norway; MNiSW and NCN, Poland; GRICES and FCT, Portugal; MNE/IFA, Romania; MES of Russia and ROSATOM, Russian Federation; JINR; MSTD,
Serbia; MSSR, Slovakia; ARRS and MIZ\v{S}, Slovenia; DST/NRF, South Africa;
MINECO, Spain; SRC and Wallenberg Foundation, Sweden; SER, SNSF and Cantons of
Bern and Geneva, Switzerland; NSC, Taiwan; TAEK, Turkey; STFC, the Royal
Society and Leverhulme Trust, United Kingdom; DOE and NSF, United States of
America.

The crucial computing support from all WLCG partners is acknowledged
gratefully, in particular from CERN and the ATLAS Tier-1 facilities at
TRIUMF (Canada), NDGF (Denmark, Norway, Sweden), CC-IN2P3 (France),
KIT/GridKA (Germany), INFN-CNAF (Italy), NL-T1 (Netherlands), PIC (Spain),
ASGC (Taiwan), RAL (UK) and BNL (USA) and in the Tier-2 facilities
worldwide.

\bibliography{paper}

\begin{thebibliography}{81}
\bibitem[{atl()}]{atlascoord}
\bibinfo{howpublished}{{ATLAS uses a right-handed coordinate system with its
  origin at the nominal interaction point (IP) in the centre of the detector
  and the $z$-axis along the beam pipe. The $x$-axis points from the IP to the
  centre of the LHC ring, and the y-axis points upwards. Cylindrical
  coordinates $(r,\phi)$ are used in the transverse plane, $\phi$ being the
  azimuthal angle around the beam pipe. The pseudorapidity is defined in terms
  of the polar angle $\theta$ as $\eta = - \ln \tan(\theta/2)$.}}
\end{thebibliography}

\onecolumngrid
\begin{flushleft}
{\Large The ATLAS Collaboration}

\bigskip

G.~Aad$^{\rm 48}$,
T.~Abajyan$^{\rm 21}$,
B.~Abbott$^{\rm 112}$,
J.~Abdallah$^{\rm 12}$,
S.~Abdel~Khalek$^{\rm 116}$,
O.~Abdinov$^{\rm 11}$,
R.~Aben$^{\rm 106}$,
B.~Abi$^{\rm 113}$,
M.~Abolins$^{\rm 89}$,
O.S.~AbouZeid$^{\rm 159}$
% ...
\bigskip
\\
% ...
$^{11}$ Institute of Physics, Azerbaijan Academy of Sciences, Baku, Azerbaijan\\
$^{12}$ Institut de F{\'\i}sica d'Altes Energies and Departament de F{\'\i}sica de la Universitat Aut{\`o}noma de Barcelona, Barcelona, Spain\\
% ...
$^{21}$ Physikalisches Institut, University of Bonn, Bonn, Germany\\
% ...
$^{48}$ Fakult{\"a}t f{\"u}r Mathematik und Physik, Albert-Ludwigs-Universit{\"a}t, Freiburg, Germany\\
% ...
$^{89}$ Department of Physics and Astronomy, Michigan State University, East Lansing MI, United States of America\\
% ...
$^{106}$ Nikhef National Institute for Subatomic Physics and University of Amsterdam, Amsterdam, Netherlands\\
% ...
$^{112}$ Homer L. Dodge Department of Physics and Astronomy, University of Oklahoma, Norman OK, United States of America\\
$^{113}$ Department of Physics, Oklahoma State University, Stillwater OK, United States of America\\
% ...
$^{116}$ LAL, Universit{\'e} Paris-Sud and CNRS/IN2P3, Orsay, France\\
% ...
$^{159}$ Department of Physics, University of Toronto, Toronto ON, Canada\\
% ...
\end{flushleft}


\end{document}